\documentclass[aps,prd,twocolumn,reprint,preprintnumbers,nofootinbib,superscriptaddress,floatfix,longbibliography]{revtex4-1}
\pdfoutput=1
\usepackage{graphicx}
\usepackage{epstopdf}
\usepackage{mathrsfs}
\usepackage{amssymb}
\usepackage{verbatim}
\usepackage{color}
\usepackage{multirow}
\usepackage{stackrel}

\usepackage{amsmath}

\DeclareMathOperator*{\argmin}{arg\,min}

\newcommand{\f}{\mathbf{f}}

\newcommand{\beq}{\begin{equation}}
\newcommand{\eeq}{\end{equation}}
\newcommand{\ga}{\lower.7ex\hbox{$\;\stackrel{\textstyle>}{\sim}\;$}}
\newcommand{\la}{\lower.7ex\hbox{$\;\stackrel{\textstyle<}{\sim}\;$}}

\usepackage{hyperref}
\usepackage{amsmath,bm}
\usepackage{physics}
\allowdisplaybreaks

\hypersetup{
    colorlinks = true,
    citecolor = {blue},
    linkcolor = {blue},
    urlcolor = {blue},
}

\begin{document}

\title{Deep Learning Symmetries and Their Lie Groups, Algebras, \\ and Subalgebras from First Principles}

\author{Roy Forestano}
\email{roy.forestano@ufl.edu}
\affiliation{Institute for Fundamental Theory, Physics Department, University of Florida, Gainesville, FL 32611, USA}

\author{Konstantin T.~Matchev}
\email{matchev@ufl.edu}
\affiliation{Institute for Fundamental Theory, Physics Department, University of Florida, Gainesville, FL 32611, USA}

\author{Katia Matcheva}
\email{matcheva@ufl.edu}
\affiliation{Institute for Fundamental Theory, Physics Department, University of Florida, Gainesville, FL 32611, USA}

\author{Alexander Roman}
\email{alexroman@ufl.edu}
\affiliation{Institute for Fundamental Theory, Physics Department, University of Florida, Gainesville, FL 32611, USA}

\author{Eyup Unlu}
\email{eyup.unlu@ufl.edu}
\affiliation{Institute for Fundamental Theory, Physics Department, University of Florida, Gainesville, FL 32611, USA}

\author{Sarunas~Verner}
\email{verner.s@ufl.edu}
\affiliation{Institute for Fundamental Theory, Physics Department, University of Florida, Gainesville, FL 32611, USA}

\begin{abstract}
We design a deep-learning algorithm for the discovery and identification of the continuous group of symmetries present in a labeled dataset. We use fully connected neural networks to model the symmetry transformations and the corresponding generators. We construct loss functions that ensure that the applied transformations are symmetries and that the corresponding set of generators forms a closed (sub)algebra. Our procedure is validated with several examples illustrating different types of conserved quantities preserved by symmetry. In the process of deriving the full set of symmetries, we analyze the complete subgroup structure of the rotation groups $SO(2)$, $SO(3)$, and $SO(4)$, and of the Lorentz group $SO(1,3)$. Other examples include squeeze mapping, piecewise discontinuous labels, and $SO(10)$, demonstrating that our method is completely general, with many possible applications in physics and data science. Our study also opens the door for using a machine learning approach in the mathematical study of Lie groups and their properties.
\end{abstract}

\date{January 12, 2023}

\maketitle
\tableofcontents

\section{Introduction}

Symmetries play a fundamental role in modern physics \cite{Gross1996}. Physical systems with continuous symmetries exhibit conservation laws that are universally applicable and indispensable in understanding the system's behavior and evolution. In particle physics, symmetries provide an organizing principle behind the observed particle zoo and its interactions, and guide model-builders in the search for viable extensions of the Standard Model (SM)\cite{Csaki:2018muy}. At the same time, the mathematical study of symmetries is interesting in its own right and has a rich history.

Over the last decade, there has been increased interest in applications of machine learning (ML) to high-dimensional physics data as a sensitive tool for event simulation, data analysis, and statistical inference \cite{Bourilkov:2019yoi,Calafiura2022,Plehn:2022ftl}. More recently, ML is also being used to facilitate tasks that traditionally have fallen within the domain of theorists, e.g., performing symbolic computations \cite{Dersy:2022bym,Alnuqaydan:2022ncd} or deriving analytical formulas by training a symbolic regression on synthetic data \cite{Choi:2010wa,Udrescu:2019mnk,Lample1912.01412,Cranmer:2020wew,Butter:2021rvz,Arechiga2112.04023,Matchev2022ApJ,Ascoli2201.04600,Lemos:2022cdj,Kamienny2204.10532,Li2205.11798,Matsubara2206.10540,Dong:2022trn}.

Applications of ML to the study of symmetries have been pursued by a number of groups in different contexts. One line of work investigates how a given symmetry is reflected in a learned representation of the data \cite{Iten1807.10300,Dillon:2021gag} or in the ML architecture itself, e.g., in the embedding layer of a neural network (NN) \cite{Krippendorf:2020gny}. Several proposals attempt to design special ML architectures (equivariant NNs) which have a desired symmetry property built in from the outset \cite{Butter:2017cot,Kanwar:2020xzo,Bogatskiy:2020tje,Gong:2022lye,Bogatskiy:2022hub,Li:2022xfc,Hao:2022zns} and test their performance \cite{Gruver2210.02984}. Incorporating the symmetry directly into the ML model makes it more economical (in terms of learned representations), interpretable and trainable. The approach can be extended to cover discrete (permutation) symmetries as well \cite{Fenton:2020woz,Shmakov:2021qdz}. Such efforts pave the way for data-driven blind searches for new physics which stress-test the data for violations of a well-established symmetry of the SM \cite{Tombs:2021wae,Lester2021,Birman:2022xzu}.

More recently, machine learning is also being applied to address more formal theoretical questions. For example, a good understanding of the symmetries present in the problem can reveal conserved quantities \cite{2003.04630,Liu:2020omw} or hint at a more fundamental unified picture \cite{Wu_2019}. ML has been used to discover the symmetry of a potential \cite{Krippendorf:2020gny,Barenboim:2021vzh,Craven:2021ems}, to decide whether a given pair of inputs is related by symmetry or not \cite{Wetzel:2020jan}, to distinguish between scale-invariant and conformal symmetries \cite{Chen:2020dxg}, and to explore the string landscape \cite{He:2017set,Carifio:2017bov,Ruehle:2020jrk}. Recent work made use of Generative Adversarial Networks to learn transformations that preserve probability distributions \cite{Desai:2021wbb}. ML applications have also found their way into group theory, which provides the abstract mathematical language of symmetries. For example, recent work used ML to compute tensor products and branching rules of irreducible representations of Lie groups \cite{Chen:2020jjw} and to obtain Lie group generators of a symmetry present in the data \cite{Liu:2021azq, Moskalev2210.04345}.

The main goal of this paper is to design a deep-learning method that mimics the traditional theorist's thinking and is capable of discovering and categorizing the full set of (continuous) symmetries in a given dataset {\em from first principles}, i.e., without any prior assumptions or prejudice. The only input to our procedure is a labeled dataset $\{\mathbf{x};y\}$ like the one in eq.~(\ref{eq:dataset}) below. An oracle $\varphi(\mathbf{x})=y$ can then be learned from the dataset, or alternatively, can be provided externally. With those ingredients, we go through the following objectives:
\\
\begin{itemize}
\item {\bf Discovery of symmetries.} In the first step, described in Section~\ref{sec:Invariance}, we learn to generate a symmetry transformation, $\mathbf{x}\stackrel{\mathbf{f}}{\to} \mathbf{x}'$, which preserves the oracle values. The transformation $\mathbf{f}$ is encoded in a neural network trained on the dataset. In general, there will be many possible symmetry transformations $\f$, and their study and categorization from a group theory point of view is the main goal of this paper. 
\item {\bf Discovery of symmetry generators.} Having learned how to generate arbitrary (finite) symmetry transformations $\mathbf{f}$, we then focus on {\em infinitesimal} transformations $\delta\mathbf{f}$, which give us in turn the symmetry {\em generators} $\mathbf{J}$. 
\item {\bf Discovery of Lie subalgebras.} By adding suitable terms to the loss function, our procedure requires that the learned set of generators $\{\mathbf{J}_\alpha\}$ forms a closed algebra. This allows us to discover subalgebras of the symmetry group, and identify them by their structure constants. Since the number of generators $N_{g}$ is a free input, in cases when the training is unsuccessful (as quantified by the loss), we can rule out the existence of $N_{g}$-dimensional subalgebras.
\item {\bf Discovery of the full Lie algebra.} The maximum value of $N_{g}$ which gives a vanishing loss, indicates the dimension of the full Lie algebra. The corresponding learned set of generators describes the full symmetry group of the dataset.
\item {\bf Identification of the symmetry group and its subgroups.} The learned sets of generators found in the previous two steps are then used to obtain the structure constants of the respective full algebra and its subalgebras, and thus to identify the corresponding symmetry group and its subgroups. 
\end{itemize}

Our study complements and extends previous related work in \cite{Krippendorf:2020gny,Barenboim:2021vzh,Liu:2021azq,Craven:2021ems,Moskalev2210.04345}. We note that our procedure is completely general, and does not anticipate what symmetries might be present in the dataset --- instead, the symmetries are learned from scratch. In addition, our method is basis-independent since we do not choose a specific convenient basis for the learned transformations and generators. Consequently, our results will not always match the nice canonical forms of the generators found in the group theory textbooks. Nevertheless, as the examples below explicitly demonstrate, all our learned transformations and generators will satisfy the defining properties of the respective symmetry groups and subgroups.

The paper is organized as follows. In Section~\ref{sec:notation} we introduce the setup for our analysis and the corresponding notation and conventions. The main steps of our deep-learning procedure are outlined in Section~\ref{sec:deep-learning}, which also explains and motivates the necessary ingredients for the loss function. For readers who are not yet fully comfortable with a deep-learning approach, Section~\ref{sec:matrix} outlines an analogous linear algebra approach that often (but not always --- see Section~\ref{sec:discontinuous} for counterexamples) can accomplish similar objectives. Each of the remaining sections contains a separate completely worked-out example illustrating our method. The examples are distinguished by the choice of oracle and the dimensionality of the feature space. In Sections~\ref{sec:rotations2dim}-\ref{sec:rotations4dim} we choose an oracle that returns the Euclidean distance in feature space, whose dimensionality, in turn, is chosen to be $n=2$ in Section~\ref{sec:rotations2dim}, $n=3$ in Section~\ref{sec:rotations3dim}, and $n=4$ in Section~\ref{sec:rotations4dim}. In Section~\ref{sec:Lorentz} we focus on the Lorentz group, i.e., the oracle computes the pseudo-Euclidean distance in four-dimensional Minkowski space-time. In Section~\ref{sec:squeeze} we consider the squeeze transformations in $n=2$ dimensions whereby the oracle returns the product of the two features. To demonstrate the universal applicability of our technique, in Section~\ref{sec:discontinuous} we show two examples of discontinuous oracles. We summarize and conclude in Section~\ref{sec:conclusions}.

\section{Setup and notations}
\label{sec:notation}

Our starting point is a labeled dataset containing $m$ samples of $n$ features and a target label $y$:
\beq
m\ {\rm samples}
\left\{
\begin{array}{ccccc}
    x_1^{(1)}, &  x_1^{(2)}, & \ldots , & x_1^{(n)}; & y_1\\
    x_2^{(1)}, &  x_2^{(2)}, & \ldots , & x_2^{(n)}; & y_2\\
    \vdots     & \vdots      & \vdots   & \vdots     & \vdots  \\
    x_m^{(1)}, &  x_m^{(2)}, & \ldots , & x_m^{(n)}; & y_m\\
\end{array}
\right.
\, .
\label{eq:dataset}
\eeq
In ML parlance, the dataset (\ref{eq:dataset}) is an $m\times n$ dataframe with $n$ features and $m$ samples. In what follows, we use the sample index a lot more often than the feature index, thus we use an explicit subscript for the sample index and hide the feature index in the boldface vector notation $\mathbf{x}$:
\beq
{\mathbf x}\equiv \{x^{(1)}, x^{(2)},\ldots, x^{(n)}  \}.
\eeq
This allows us to write the input features in a compact form as
\beq
\left\{\mathbf{x}_i\right\} \; \equiv \; \left\{\mathbf{x}_1,\mathbf{x}_2,\ldots,\mathbf{x}_m\right\} \, .
\label{eq:set}
\eeq

In order to study the symmetries of the data (\ref{eq:dataset}), we need to know the function $y(\mathbf{x})$, which can be given analytically or numerically in terms of an oracle $\varphi: \mathbb{R}^n \rightarrow{\mathbb{R}}$ capable of computing the corresponding output target labels $y_1,y_2,\ldots,y_m$:
\beq
y_i \; = \; \varphi(\mathbf{x}_i) \, , \qquad i \, = \,1,2,\ldots,m \, .
\label{eq:labels}
\eeq
This leads to two basic scenarios:
\begin{itemize}
\item The function $y(\mathbf{x})$ is known analytically, and that same function has been used as in (\ref{eq:labels}) to compute the labels in (\ref{eq:dataset}) exactly. This case is of interest to theorists, and this is the approach adopted in this paper as well --- for each of our examples below, we specify the relevant analytic oracle $\varphi(\mathbf{x})$ and proceed to study the resulting symmetries. Note, however, that we never take advantage of the knowledge of the analytical form of the oracle, e.g., we do not differentiate or symbolically manipulate in any other way the function $\varphi(\mathbf{x})$. For our purposes, we only use the oracle numerically --- our method treats it simply as a black box, which, given the values of $\mathbf{x}$, can produce the numerical value of the label $y$.
\item The functional dependence $y(\mathbf{x})$ is a priori unknown and the dataset (\ref{eq:dataset}) is all that is available to us. This is the typical case encountered by data scientists. Now, one needs to go through a preliminary step of first creating the oracle $\varphi$ (usually in the form of a neural network trained on the dataset (\ref{eq:dataset})), which is capable of computing and reporting the values of $y=\varphi(\mathbf x)$ to us. This is a standard regression task which is of no interest here since it can be accomplished using one of the many established ML regression techniques. We can therefore safely assume that this preliminary step has already been completed and we have such a numerical oracle $y=\varphi(\mathbf{x})$ already available. 
\end{itemize}
Since we are only using the oracle numerically, from our point of view there is no real difference between the above two cases. In what follows the oracle $\varphi(\mathbf{x})$ will be used in the exact same way, regardless of its origin.

Given this general setup, our main task is to derive the symmetry transformation 
$\f: \mathbb{R}^n \rightarrow \mathbb{R}^n$
\beq
\mathbf{x}' \; = \; \f(\mathbf{x})  \, ,
\label{eq:symmetry}
\eeq
which preserves the $\varphi$-induced labels of our dataset (\ref{eq:dataset}). In other words, we want to find the function $\f(\mathbf{x})$ for which 
\beq
\varphi(\mathbf{x}'_i) \; \equiv \; \varphi(\f(\mathbf{x}_i)) = \varphi(\mathbf{x}_i), \quad \forall i \,= \,1,2,\ldots,m \, .
\eeq
The particular instantiation of the symmetry $\f(\mathbf{x})$ will depend on the initialization of our parameters, so by repeating the procedure with different initializations, we will in principle obtain a whole family of symmetry transformations.

Next, we focus on infinitesimal symmetry transformations and proceed to study the corresponding set of generators $\{\mathbf{J}_\alpha\}$, with $\alpha=1,2,\ldots,N_g$, where we use lowercase Greek letters to label the generators of symmetry transformations. A given set of generators $\{\mathbf{J}_\alpha\}$ forms a Lie algebra if the closure condition is satisfied, i.e., if all Lie brackets $\bigl[ \, .\, ,\, .\, \bigr]$ can be represented as linear combinations of the generators in the set:
\beq
\bigl[ \mathbf{J}_\alpha, \mathbf{J}_\beta\bigr] 
= \sum_{\gamma=1}^{N_g} a_{[\alpha\beta]\gamma} \mathbf{J}_\gamma .
\label{eq:algebraclosure}
\eeq
The coefficients $a_{[\alpha\beta]\gamma}$ are the structure constants (anti-symmetric in their first two indices, as implied by the square brackets) whose values will reveal the symmetry group present in our dataset.

In principle, the number of generators $N_g$ is a hyperparameter that must be specified ahead of time (similarly to the choice of the number of clusters in certain clustering algorithms like K-means). Therefore, when we find a closed algebra at a given $N_g$ value, we are only guaranteed that it is a subalgebra, and we must proceed to try out higher values for $N_g$ as well. The full algebra will then correspond to the maximum value of $N_g$ for which a closed algebra of distinct generators is found to exist.

\section{Deep Learning Approach}
\label{sec:deep-learning}

In our approach, we model the output function $\f$ with a neural network (NN) ${\mathcal F}_{\mathcal W} $ with $n$ neurons in the output layer, corresponding to the $n$ transformed features of the data point $\mathbf{x}'$. The trainable network parameters (weights and biases) will be generically denoted with $\mathcal W$. During training, they will evolve and converge to the corresponding {\em trained} values $\widehat{\mathcal W}$ of the parameters of the trained network ${\mathcal F}_{\widehat{\mathcal W}}$, i.e., the hat symbol will denote the result of the training. Once the parameters $\widehat{\mathcal W}$ are found, we can find the structure constants using standard methods. We choose a suitable loss function that ensures the desired properties of the trained network. The following subsections discuss the individual contributions to the loss function, which in our implementation are combined and minimized simultaneously.

The neural network ${\mathcal F}_{\mathcal W}$ is implemented as a sequential feed-forward neural network in {\sc PyTorch} \cite{NEURIPS2019_9015}. 
 The examples in Sections~\ref{sec:rotations2dim}-\ref{sec:Lorentz} are simple enough to be done with no hidden layers, no bias, and no activation function, i.e., with linear transformations (see Section~\ref{sec:matrix}). For the examples in the later sections, we do add hidden layers. Optimizations are performed with the {\sc Adam} optimizer with a learning rate between $0.03$ and $0.1$. The loss functions were designed to achieve a fast and efficient training process without the need for extensive hyperparameter tuning. The training data (\ref{eq:set}) was typically on the order of a few hundred points and was sampled from a standard normal distribution. 

An alternative approach to predicting the generators from the model parameters, which utilizes an identical loss function, can be carried out where a NN, ${\mathcal F}: \{\mathcal{G}_\alpha, \mathbf{a}  \} \to \{\widehat{\mathcal G}_\alpha, \hat{\mathbf{a}} \}$, takes a set of randomly initialized $n \times n$ generators $\{\mathcal{G}_\alpha \}$ and an $N_b \times N_g$ structure constant array $\{ \mathbf{a} \}$, flattens each individual generator and the structure constant array, and proceeds to converge the elements to the desired set of generators $\{\widehat{\mathcal{G}_\alpha} \}$ and structure constant array $\{ \hat{\mathbf{a}}\}$. Here $N_b = {N_g \choose 2}  = \frac{N_g(N_g-1)}{2}$ denotes the number of unique one-directional Lie brackets for a given number of generators $N_g$.

 This alternative approach can be implemented as a module list of sequential neural network layers consisting of two hidden layers for the generators and a single sequential layer consisting of two hidden layers for the structure constants. Each layer consists of a bias and the hidden layers use the Rectified Linear Unit (ReLU) activation function. The optimizer, average learning rate, model hyperparameters, and training data generation were identical to the alternative implementation described above. Whereas the original method feeds the data directly into the network, the data in this approach is fed into the loss function to be transformed by the model's predicted generators.

\subsection{Invariance}
\label{sec:Invariance}

The invariance under the transformation (\ref{eq:symmetry}) is enforced by requiring that the labels before and after the transformation remain the same. For this purpose, we include the following mean squared error (MSE) term in the loss function $L$:
\beq
    L_{inv}(\mathcal W, \{\mathbf x_i\}) = \frac{1}{m}\sum_{i=1}^m \left[ \varphi\left({\mathcal F}_{\mathcal W}(\mathbf{x}_i)\right)-\varphi(\mathbf{x}_i) \right]^2 \, .
\label{eq:lossInv}
\eeq
A NN trained with this loss function will find an arbitrarily general (finite) symmetry transformation ${\mathcal F}_{\widehat{\mathcal W}}$ parametrized by the values of the trained network parameters $\widehat{\mathcal W}$.

In order to find multiple symmetries, the process can be repeated. Alternatively, several networks can be trained concurrently, by modifying the loss function to ensure that the resulting transformations are sufficiently distinct (see Section~\ref{sec:orthogonality} below). 

\subsection{Infinitesimality}

In order to focus on the {\em generators} of the possible set of symmetry transformations, we restrict ourselves to infinitesimal transformations $\delta{\mathcal F}$ in the vicinity of the identity transformation $\mathbf{I}$:
\beq
\delta{\mathcal F} \; \equiv \; \mathbf{I} + \varepsilon \, {\mathcal G}_{\mathcal W} \, ,
\label{eq:fepsilon}
\eeq
where $\varepsilon$ is an infinitesimal parameter and the parameters ${\mathcal W}$ of the new neural network ${\mathcal G}$ will be forced to be finite, which ensures that (\ref{eq:fepsilon}) is an infinitesimal transformation. The loss function (\ref{eq:lossInv}) can then be rewritten as
\beq
    L_{inf}(\mathcal W, \{\mathbf x_i\}) = \frac{1}{m\varepsilon^2}\sum_{i=1}^m \left[ \varphi(\mathbf{x}_i + \varepsilon {\mathcal G}_{ \mathcal W}(\mathbf{x}_i))-\varphi(\mathbf{x}_i) \right]^2 \, ,
\label{eq:LossInvInfinitesimal}    
\eeq
where we have introduced an explicit factor of $\varepsilon^2$ in the denominator to account for the fact that generic transformations scale as $\varepsilon$ \cite{Craven:2021ems}. Once we minimize the loss function, the trained NN ${\mathcal G}_{\widehat {\mathcal W}}$ will represent a corresponding generator
\beq
\mathbf{J} \; = \; {\mathcal G}_{\widehat {\mathcal W}} \, ,
\eeq
where
\beq
\widehat{\mathcal W}\equiv\argmin_{\mathcal W}
\biggl(L({\mathcal W}, \{\mathbf x_i\}) \biggr)
\eeq
are the learned values of the NN parameters which minimize the loss function. The result for $\widehat{\mathcal W}$, and therefore, for $\mathbf{J}$, will in principle depend on the starting values ${\mathcal W}_0$ of the network parameters at initialization. If we now repeat the training $N_g$ times under different initial conditions ${\mathcal W}_0$ (or with different values of the hyperparameters), we will obtain a set of $N_g$ (in general distinct) generators $\{\mathbf{J}_\alpha\}$, $\alpha=1,2,\ldots,N_g$.

\subsection{Orthogonality}
\label{sec:orthogonality}

In order to make the set of generators $\{\mathbf{J}_\alpha\}$, $\alpha=1,2,\ldots,N_g$, found in the previous step maximally different, we introduce the following additional orthogonality term to the loss function
\beq
 L_{ortho}({\mathcal W}, \{\mathbf x_i\}) = 
\sum_{ \alpha < \beta}^{N_g} 
\left(
\sum_{p}
{\mathcal W}_\alpha^{(p)} {\mathcal W}_\beta^{(p)}\right)^2, 
\label{eq:lossOrtho}
\eeq
where the index $p$ runs over the individual NN parameters ${\mathcal W}_\alpha^{(p)}$. 

\subsection{Closure}

In order to test whether a certain set of {\em distinct} generators $\{\mathbf{J} _\alpha\}$, $\alpha=1,2,\ldots,N_g$, found in the previous steps, generates a group, we need to check the closure of the algebra (\ref{eq:algebraclosure}). This can be done in several ways. The most principled would be to minimize
\beq
L_{closure} (a_{[\alpha\beta]\gamma}) = \sum_{\alpha<\beta}\Tr \left(\mathbf{C}_{[\alpha\beta]}  ^T\mathbf{C}_{[\alpha\beta]}\right),
\label{eq:LossClosure}
\eeq
with respect to the candidate structure constant parameters $a_{[\alpha\beta]\gamma}$, where the closure mismatch is defined by
\beq 
\mathbf{C}_{[\alpha\beta]} (a_{[\alpha\beta]\gamma}) \equiv
\bigl[ \mathbf{J}_\alpha, \mathbf{J}_\beta\bigr] 
- \sum_{\gamma=1}^{N_g} a_{[\alpha\beta]\gamma} \mathbf{J}_\gamma .
\label{eq:closuremismatch}
\eeq
Since $L_{closure}$ is positive semi-definite, $L_{closure}=0$ would indicate that the algebra is closed and we are thus dealing with a genuine (sub)group. In practice,\footnote{Another possible approach is to minimize the out-of-space components of the commutators with respect to the space of generators, after flattening and Gram–Schmidt orthonormalization.} we perform the minimization of (\ref{eq:LossClosure}) simultaneously with the previously discussed contributions to the loss function, by replacing ${\mathbf J}_\alpha \to {\mathcal G}_{\mathcal W_\alpha}$ in (\ref{eq:closuremismatch}). The advantage of this simultaneous construction is that  every set of generators that we obtain at any given value of $N_g$ is already forming an algebra that is ``as closed as possible". Then, the size of the achieved total training loss is an indicator whether for that value of $N_g$ a closed algebra exists or not.

Once a closed set of valid generators has been found, we can retrain the NN in a conveniently chosen canonical basis and obtain the canonical form of the set of generators, whose structure constants in turn reveal the nature of the group behind the found symmetry transformations ${\mathcal G}_{\widehat{\mathcal W}_\alpha}$, $\alpha=1,2,\ldots,N_g$.

Sections~\ref{sec:rotations2dim}-\ref{sec:discontinuous} illustrate the steps above with several examples of increasing complexity. 

\section{Linear Algebra Approach}
\label{sec:matrix}

The universal approximation theorems \cite{HORNIK1989359} guarantee that the deep-learning approach of the previous section can handle almost any symmetry transformation, including a highly non-linear one. At the same time, a large class of interesting symmetries arising in physics are {\em linear} transformations for which the usual formalism of linear algebra would suffice. Furthermore, the analysis of the symmetry generators involves infinitesimal transformations, which are represented with linear operators. For those reasons and to captivate the readers who are not yet at ease with the technical intricacies of machine learning, in this section, we reformulate our analysis from the previous section in the language of linear algebra. We follow the same notation and conventions, but replace the calligraphic font symbols representing neural networks with corresponding blackboard-bold symbols representing $n\times n$ matrices acting on the $n$-dimensional feature space.

In this section we are interested in the linear subclass of the transformations (\ref{eq:symmetry}), which are encoded in a generic matrix $\mathbb{F}$
\beq
\mathbf{x}' = \mathbb{F}\, \mathbf{x} \, ,
\eeq
whose $n^2$ components $\mathbb{F}_{ij}$ are determined by minimizing the loss function
\beq
    L_{inv}(\mathbb F, \{\mathbf x_i\}) = \frac{1}{m}\sum_{i=1}^m \left[ \varphi\left({\mathbb F}\,\mathbf{x}_i\right)-\varphi(\mathbf{x}_i) \right]^2 \, .
\label{eq:lossInvMatrix}
\eeq
If the minimization is successful, then such a linear symmetry exists and is represented with the learned matrix $\widehat{\mathbb F}$.

In analogy to (\ref{eq:fepsilon}), we can write the corresponding infinitesimal linear transformation as
\begin{equation}
    \delta \mathbb{F}(\varepsilon) = \mathbb{I} + \varepsilon\, \mathbb{G} \, ,
    \label{eq:rotinf2}
\end{equation}
where $\mathbb{I}$ is the unit $n\times n$ matrix and $\mathbb{G}$ is an $n\times n$ matrix whose components $\mathbb{G}_{ij}$ are yet to be determined through the optimization. In order to obtain a {\em single} generator matrix $\mathbb{J}$, we can use a loss function
analogous to (\ref{eq:LossInvInfinitesimal})
\begin{eqnarray}
    L_{\mathbb{J}} (\mathbb{G},\{\mathbf{x}_i\}) &=& \frac{h_{inv}}{m\varepsilon^2} \sum_i^m\biggl[\varphi\bigl(
    \mathbf{x}_i+\varepsilon\,\mathbb{G}\, \mathbf{x}_i
    \bigr) - \varphi(\mathbf{x}_i) \biggr]^2  \nonumber \\
    &+& h_{norm} \biggl[\Tr (\mathbb G^T \mathbb{G}) - 2\biggr]^2 ,
    \label{eq:loss_si}
\end{eqnarray}
where the constants $h_{inv}$ and $h_{norm}$ in (\ref{eq:loss_si}) are hyperparameter weights determining the relative contribution of the two terms in the loss function (\ref{eq:loss_si}) enforcing the symmetry invariance and finite normalization\footnote{In order for (\ref{eq:rotinf2}) to be an infinitesimal transformation, the matrix $\mathbb G$ needs to be finite. We choose to normalize our generators as ${\rm Tr}({\mathbb G}_\alpha^T {\mathbb G}_\beta)=2\delta_{\alpha\beta}$, hence the factor of 2 in the second line of (\ref{eq:loss_si}).} conditions, respectively. The actual generator ${\mathbb{J}}$ is then obtained by minimizing (\ref{eq:loss_si}):
\beq
\mathbb{J} = \argmin_\mathbb{G} \biggl(  L_{\mathbb{J}} (\mathbb{G},\{\mathbf{x}_i\})\biggr) \, .
\eeq

By repeating this procedure several times with different initial conditions, we obtain a different generator $\mathbb{J}$ each time. Alternatively, we can produce all $N_g$ generators in one go by adding together and minimizing simultaneously $N_g$ copies of the loss function (\ref{eq:loss_si}):
\beq
\sum_{\alpha=1}^{N_g} L_{\mathbb{J}} (\mathbb{G}_\alpha,\{\mathbf{x}_i\})\, ,
\label{eq:loss_si_R}
\eeq
which will lead to a set of $N_g$ generators $\mathbb{J}_\alpha$, $\alpha=1,2,\ldots, N_g$, and their respective infinitesimal transformations $\delta\mathbb{F}_\alpha \equiv \mathbb{I} + \varepsilon\, \mathbb{J}_\alpha$. 

At this point the generators $\mathbb{J}_\alpha$ are completely decoupled and independent of each other. We can force them to be different by adding a loss term analogous to (\ref{eq:lossOrtho}):
\beq
 L_{ortho}({\mathbb G}, \{\mathbf x_i\}) = 
 \sum_{ \alpha < \beta}^{N_g} \left[ \text{Tr}({\mathbb G}_\alpha^T {\mathbb G}_\beta)\right]^2.
\label{eq:lossOrthoMatrix}
\eeq

Finally, we can enforce the closure property by including a loss term analogous to (\ref{eq:LossClosure}):
\beq
L_{closure} (a_{[\alpha\beta]\gamma}) = \sum_{\alpha<\beta}\Tr \left(\mathbb{C}_{[\alpha\beta]}  ^T\mathbb{C}_{[\alpha\beta]}\right),
\label{eq:LossClosureMatrix}
\eeq
where 
\beq 
\mathbb{C}_{[\alpha\beta]} (a_{[\alpha\beta]\gamma}) \equiv
\bigl[ \mathbb{G}_\alpha, \mathbb{G}_\beta\bigr] 
- \sum_{\gamma=1}^{N_g} a_{[\alpha\beta]\gamma} \mathbb{G}_\gamma \, ,
\eeq
and minimizing the total loss function with respect to the parameters $a_{[\alpha\beta]\gamma}$ as well.

\section{Length-preserving Transformations in Two Dimensions}
\label{sec:rotations2dim}

\begin{figure}[t]
    \centering
    \includegraphics[width=0.45\textwidth]{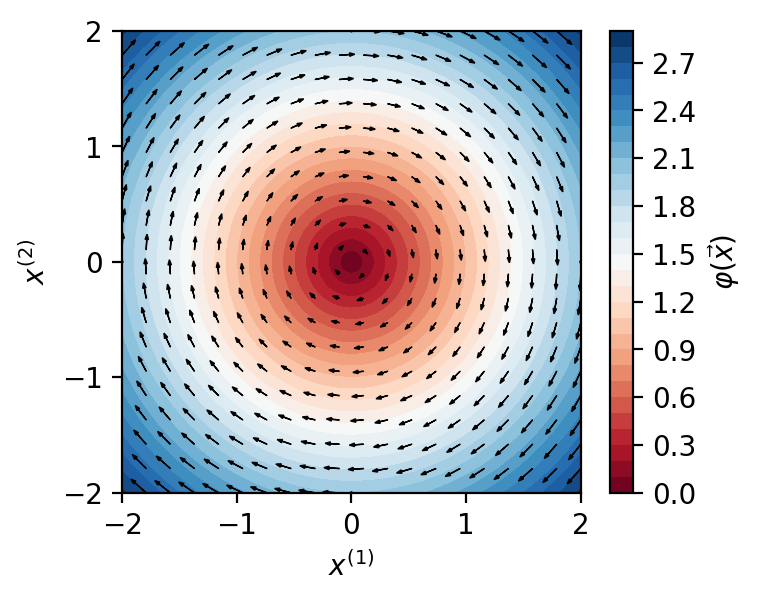}
    \caption{A representative symmetry transformation preserving the oracle function (\ref{eq:length}), found by our procedure in the two-dimensional exercise in Section~\ref{sec:rotations2dim}. In this and all subsequent such figures, the arrows represent the displacements ${\mathbf x}'-{\mathbf x}$ resulting from the symmetry transformation. The color map illustrates the oracle function $\varphi(\mathbf x)$.  } \label{fig:rotation_n2}
\end{figure}

In this and the next two sections, we focus on transformations which preserve the Euclidean length of the feature vector, i.e., our oracle $\varphi$ will return 
\beq
\varphi(\mathbf{x}) \equiv |\mathbf{x}| = \left\{\sum_{i=1}^n [x^{(i)}]^2 \right\}^{1/2}.
\label{eq:length}
\eeq
In this case we expect our method to discover the symmetry of the orthogonal group $O(n)$, whose generators can be written in terms of Kronecker deltas as
\begin{equation}
    \label{eq:groupgenerator}
    \left({\mathbb O}_{mn}\right)^{ij} \; = \; \, \delta_m^i \delta_n^j - \delta_m^j \delta_n^i \, ,
\end{equation}
in which case the algebra is given by
\begin{eqnarray}
\left[{\mathbb O}_{mn}, {\mathbb O}_{pq} \right] & = &
  \delta_{np} {\mathbb O}_{mq} 
+ \delta_{mq} {\mathbb O}_{np} \nonumber \\
&-& \delta_{mp} {\mathbb O}_{nq} 
- \delta_{nq} {\mathbb O}_{mp}.
\label{eq:rotcommutator}
\end{eqnarray}
Note that the generators ${\mathbb O}_{mn}$ are labeled by two indices, which indicate the plane of rotation.

In this section we start with the simplest case of two dimensions, $n=2$, which should correspond to the single-generator group $O(2)$. First we try to generate a single generic (i.e., not necessarily infinitesimal) symmetry transformation. For this purpose, we train our network ${\mathcal F}_{\mathcal W}$ with the invariance loss (\ref{eq:lossInv}). This exercise was successful and, depending on the initialization, we found various symmetry transformations. They all involved a rotation around the origin in the $(x^{(1)}, x^{(2)})$ plane. One representative symmetry transformation is depicted in Figure~\ref{fig:rotation_n2}.

\begin{figure}[t]
    \centering
 \includegraphics[width=0.48\textwidth]{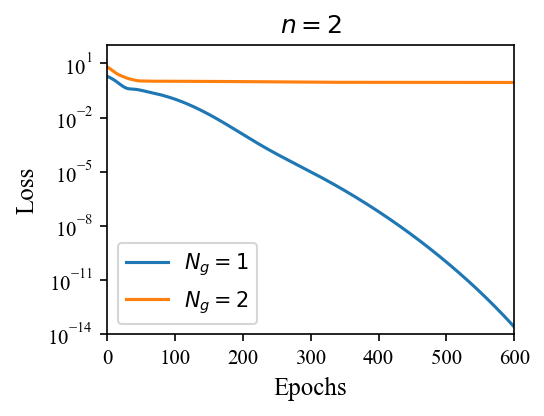}
    \caption{The evolution of the training loss with the number of epochs for the two-dimensional exercise in Section~\ref{sec:rotations2dim}. Our algorithm finds one symmetry generator (the training loss shown in blue steadily decreases) but not two different symmetry generators (the training loss shown in orange stays large). }\label{fig:lossn2}
\end{figure}

\begin{figure}[t]
    \centering
    \includegraphics[height=0.2\textwidth]{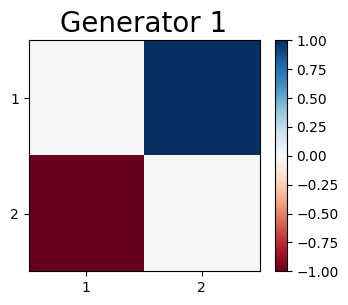} \\
    \includegraphics[height=0.2\textwidth]{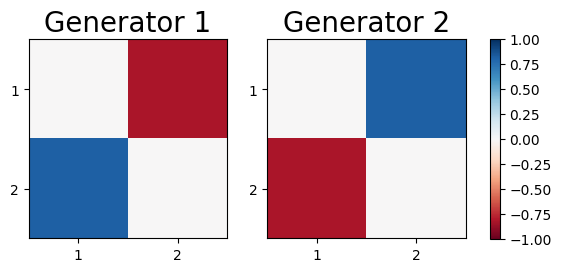}
    \caption{Top row: a successfully learned generator for $n=2$ and $N_g=1$. Bottom row: unsuccessfully learned ``generators" for $n=2$ and $N_g=2$. In this and all subsequent such figures, each panel represents a learned generator ${\mathbb J}_\alpha$ in matrix form, where the values of the individual elements of the matrix are indicated by the color bar. } \label{fig:gentsn2}
\end{figure}

Next, we turn to the algebra of symmetry generators. We begin with a single generator, $N_g=1$, in which case we do not need to include the orthogonality and closure terms in the loss function. The training is successful and the loss function is driven to zero, as seen by the blue solid line in Figure~\ref{fig:lossn2}. The resulting generator in matrix form is pictorially visualized in the top row of Figure~\ref{fig:gentsn2}, and we immediately recognize the familiar matrix ${\mathbb O}_{12}$ from (\ref{eq:groupgenerator}) generating rotations in the $12$-plane
\beq
{\mathbb O}_{12} = 
\left(
\begin{array}{rr}
0 & 1 \\
-1& 0
\end{array}
\right) \, .
\label{eq:SO2generatorMatrix}
\eeq
Note that the generator has the expected antisymmetric property.

Having found one symmetry generator, we next check if there is a second distinct generator. For this purpose, we add the orthogonality and closure terms in the loss function and repeat the training. This time the training is unsuccessful and the loss flattens after about 50 epochs, as seen by the orange solid line in Figure~\ref{fig:lossn2}. In the second row of Figure~\ref{fig:gentsn2} we show the result for the two candidate generators found in this case. We note that while they do have the expected form for a generator of rotations in two dimensions, they are essentially the same transformation, and differ only by an overall sign. This implies that they fail the orthogonality condition --- indeed, we find that the dominant contribution to the large total loss in that case is from the orthogonality loss (\ref{eq:lossOrtho}). Since the total loss is large and does not improve with further training  (see the orange line for $N_g=2$ in Figure~\ref{fig:lossn2}), these two are not valid generators and should be discarded. We thus conclude that there is no Lie algebra with $N_g=2$ distinct generators in this scenario.

We note in passing that upon repeated training runs in the $N_g=2$ case, we sometimes find that the algorithm chooses to create generators which are orthogonal to each other, but are not genuine symmetry transformations, i.e., the orthogonality loss is driven to zero, but only at the expense of a large invariance loss (\ref{eq:LossInvInfinitesimal}). In principle, it is difficult to predict what the machine will choose to do when presented with two mutually exclusive requirements. For example, in some runs the two candidate generators ended up being identical instead of differing by a minus sign. Also note that the normalization of the two candidate generators in the second row of Figure~\ref{fig:gentsn2} is off --- this is because the program chose to violate the normalization condition as well, in order to help the orthogonality loss, which is the dominant penalty in that case. In any case, since the failed training examples are of little theoretical value, in the following we shall only show results for the successful training runs which led to valid (sub)algebras.

Next, we check the cases with $N_g>2$. Similarly, we find that the training ends up in a large loss and that there is no consistent solution for a closed algebra. We therefore conclude that  in this $n=2$ example there is only a one-parameter symmetry group with a generator given by (\ref{eq:SO2generatorMatrix}).

Although the example of this section was rather trivial, it did outline and validate the main steps of our method. More complicated examples follow in the next sections.

\section{Length-preserving Transformations in Three Dimensions}
\label{sec:rotations3dim}

\begin{figure}[t]
\centering    
\includegraphics[width=0.22\textwidth]{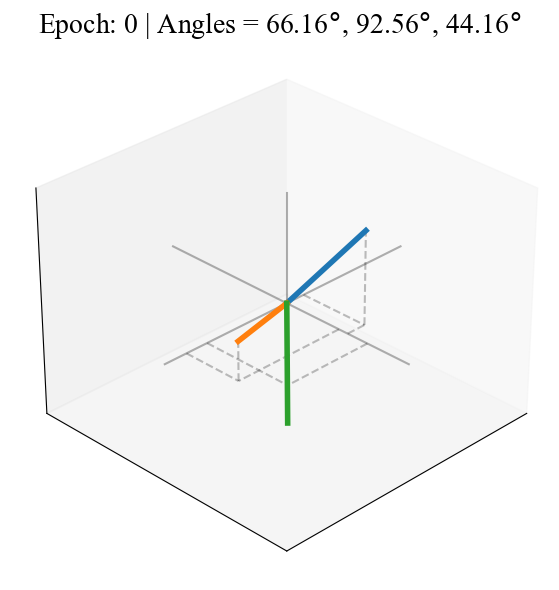}
~~
\includegraphics[width=0.22\textwidth]{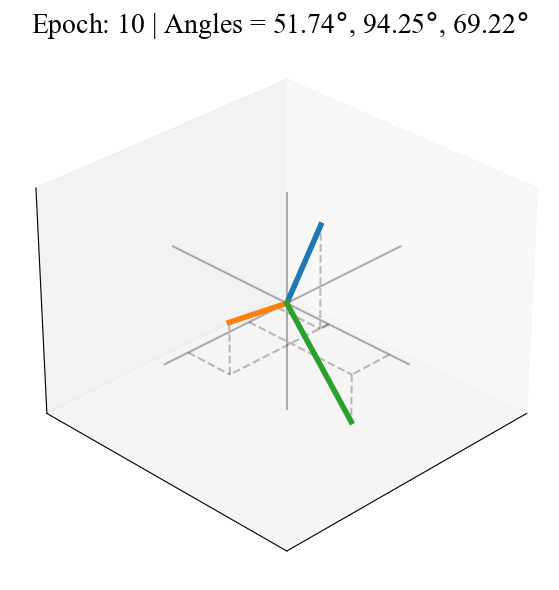}
\\
\includegraphics[width=0.22\textwidth]{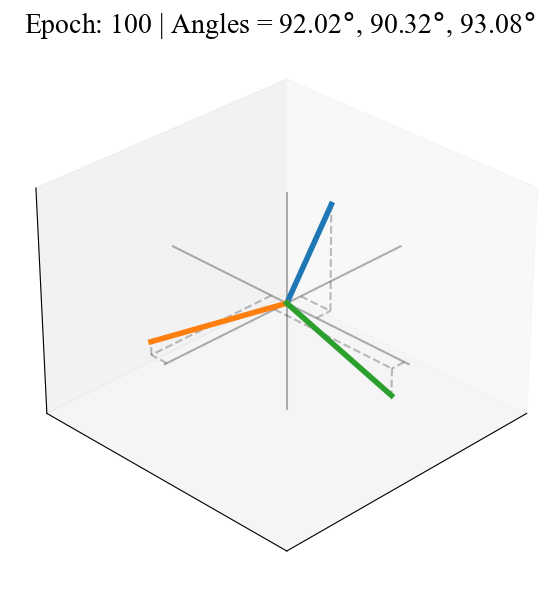}
~~
\includegraphics[width=0.22\textwidth]{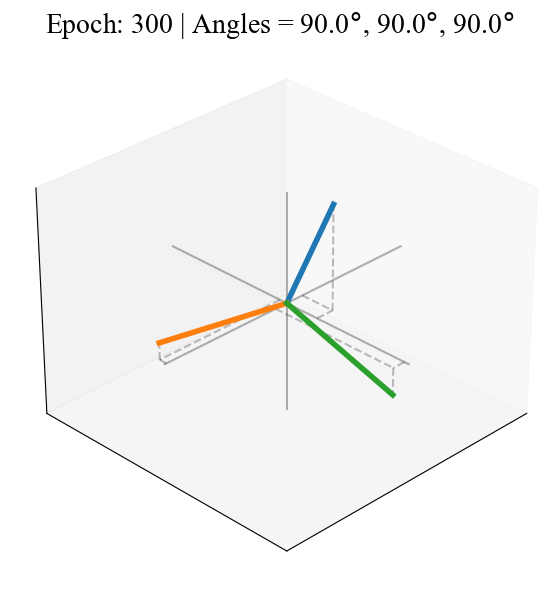}
\caption{The axes of the eigenvectors of the three generators found at intermediate stages of the training: after 1 epoch (top left), after 10 epochs (top right), after 100 epochs (bottom left), and after 300 epochs (bottom right). For convenience, at the top of each panel we list the angle (in degrees) between each pair of axes. }\label{fig:training}
\end{figure}

In this section, we proceed to study symmetry transformations which preserve the oracle (\ref{eq:length}) in $n=3$ dimensions. Once again, we find that training with the invariance loss (\ref{eq:lossInv}) alone always leads to a valid symmetry transformation. We also observe that the matrix form $\mathbb{F}$ is antisymmetric, in agreement with the expectation for the orthogonal group $O(3)$. Now that the data is three-dimensional, however, it is difficult to visualize the symmetry transformation directly as in Figure~\ref{fig:rotation_n2}. Instead, we choose to plot the symmetry axis (in this case the axis of rotation) defined by the real eigenvector. Figure~\ref{fig:training} illustrates the transformations found at different stages of the training for the case of $N_g=3$ generators. At the top of each panel we list the relative angle in degrees for each pair of axes. Note how the symmetry axes start out oriented at random, but the orthogonality loss term (\ref{eq:lossOrtho}) gradually drives them to a mutually orthogonal configuration.

\begin{figure}[t]
\centering
\includegraphics[width=0.48\textwidth]{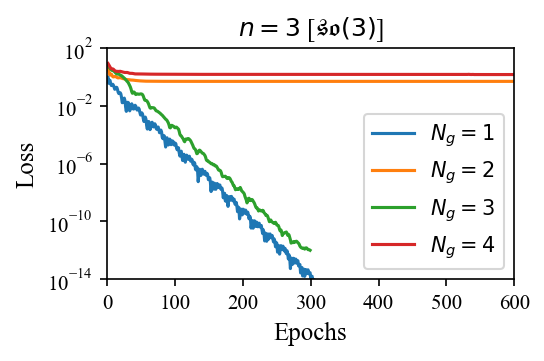}
\caption{The same as Fig.~\ref{fig:lossn2}, but for the $n=3$ exercise considered in Section~\ref{sec:rotations3dim}. The training results in valid closed algebras with one or three generators, but not two or four.}\label{fig:lossn3}
\end{figure}

In order to analyze the group structure of the found symmetry transformations, we proceed to study the generators of infinitesimal transformations. First we try to determine the dimensionality of the full algebra, i.e., the maximal number of linearly independent generators resulting in a closed algebra. Figure~\ref{fig:lossn3} shows loss curves for several different values of $N_g$: 1, 2, 3 and 4 (we do not show results for $N_g\ge 5$ since they had large losses). We observe that the training was successful only for the cases of $N_g=1$ and $N_g=3$. This implies that the full algebra has 3 generators, in agreement with the expectation of $n(n-1)/2$ generators for an orthogonal $O(n)$ group. At the same time, the successful training at $N_g=1$ reveals a single generator subgroup of $O(3)$ which is nothing but the $O(2)$ discussed in the previous section.
 
\begin{figure}[t]
    \centering
    \includegraphics[width=0.48\textwidth]{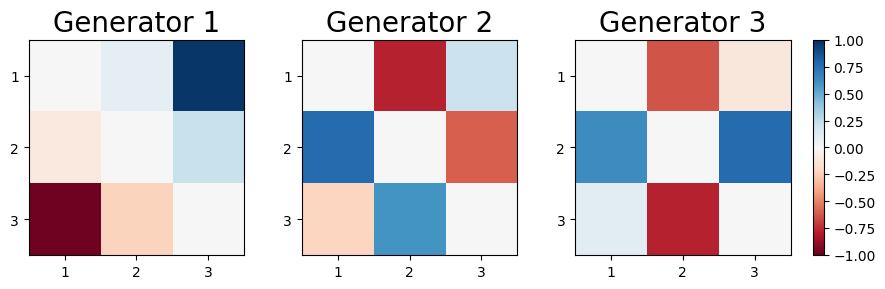} \\
    \includegraphics[width=0.3\textwidth]{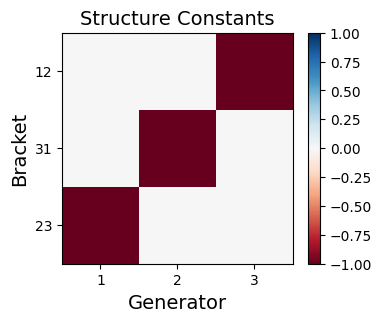}
    \caption{Top row: the successfully learned generators for the $n=3$, $N_g=3$ exercise considered in Section~\ref{sec:rotations3dim}. Bottom panel: the corresponding structure constants in matrix form (see text). }\label{fig:so3_3generators}
\end{figure}

The results from a typical training run at $N_g=3$ are shown in Figure~\ref{fig:so3_3generators}, where the top row depicts the three successfully learned generators ${\mathbb J}_\alpha$, $\alpha=1,2,3$, while the bottom panel is a pictorial representation of the structure constants in matrix form as follows. Each row (labeled $\alpha\beta=12,31,23$) represents one of the three possible commutators, whereas the columns (labeled $\gamma=1,2,3$) represent the found generators $\mathbb J_\gamma$ shown in the top panels of the figure. Then, the entry in each cell represents the structure constant $a_{[\alpha\beta]\gamma}$ from the defining equation~(\ref{eq:algebraclosure}). 

A careful inspection of the top row in Figure~\ref{fig:so3_3generators} reveals that the three generators found in our example are approximately
\begin{subequations}
\begin{eqnarray}
\mathbb J_1 &\approx & 
\left(
\begin{array}{rrr}
0 & 0 & 1 \\
0 & 0 &  0 \\
-1 & 0 &  0 
\end{array}
\right)
= -\, \mathbb O_{31},
\\
\mathbb J_2 &\approx & 
\frac{1}{\sqrt{2}}
\left(
\begin{array}{rrr}
0 & -1 &  0 \\
1 & 0 &  -1 \\
0 & 1 &  0 
\end{array}
\right)
= -\,\frac{1}{\sqrt{2}}\,
(\mathbb O_{12}+\mathbb O_{23}),~~
\\
\mathbb J_3 &\approx & 
\frac{1}{\sqrt{2}}
\left(
\begin{array}{rrr}
0 & -1 &  0 \\
1 & 0 &  1 \\
0 & -1 &  0 
\end{array}
\right)
= -\,\frac{1}{\sqrt{2}}\,
(\mathbb O_{12}-\mathbb O_{23}).
\end{eqnarray}
\end{subequations}
The bottom panel shows that the algebra of the found three generators $\mathbb J_\alpha$ is given by
\beq
\left[ \mathbb J_\alpha, \mathbb J_\beta \right] = -\,\epsilon_{\alpha\beta\gamma}   \mathbb J_\gamma ,
\label{eq:SO3algebra}
\eeq
in which we recognize the usual SO(3) algebra involving the Levi-Civita permutation symbol  $\epsilon_{\alpha\beta\gamma}$. From now on we will be using the Einstein summation convention for repeated generator-type indices $\alpha, \beta, \gamma, \dots$.

\section{Length-preserving Transformations in Four Dimensions}
\label{sec:rotations4dim}

In this section, we generalize the discussion from the previous two sections to the case of $n=4$ dimensions. The new twist here will be the existence of multiple nontrivial subalgebras of different dimensionality.

\begin{figure}[t]
    \centering
    \includegraphics[width=0.47\textwidth]{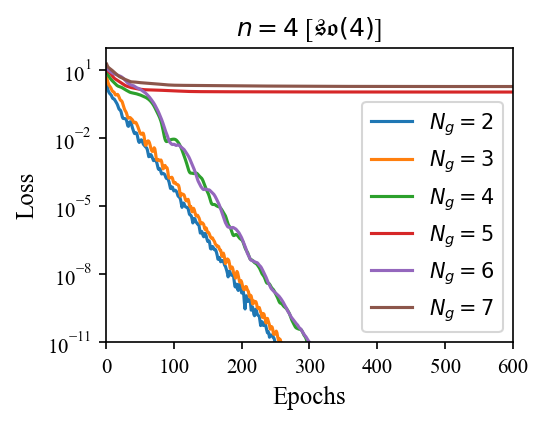}
    \caption{
    The same as Fig.~\ref{fig:lossn3}, but for the $n=4$ exercise considered in Section~\ref{sec:rotations4dim}. The training results in valid closed algebras with one (not shown), two, three, four and six generators, but not five or seven.}\label{fig:lossn4}
\end{figure}

The discovery of a single finite symmetry transformation with the invariance loss (\ref{eq:lossInv}) is straightforward and always succeeds in finding some finite rotation in four dimensions, represented with an orthogonal matrix. Therefore in this section, we focus only on the identification of the (sub)algebras. As before, we repeat the training for various number of multiple distinct generators ($N_g=2,3,4,5,6,7$) and with the orthogonality and closure losses turned on. Representative loss curves are shown in Figure~\ref{fig:lossn4}. We observe that a closed algebra is found in four of those cases, namely $N_g=2,3,4,6$, which we now discuss in turn (the trivial case of $N_g=1$ is of course always possible and will not be specifically discussed from now on).

\subsection{Two generator subalgebras}
\label{sec:Rot_n4_g2}

\begin{figure}[t]
    \centering
    \includegraphics[width=0.47\textwidth]{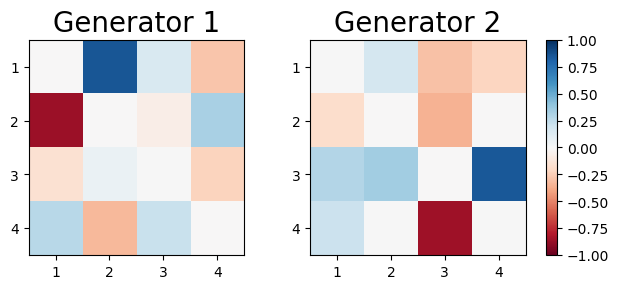}\\
    \includegraphics[width=0.3\textwidth]{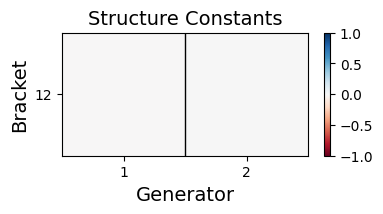}    
    \caption{Top row: the successfully learned generators for the $n=4$, $N_g=2$ exercise considered in Section \ref{sec:Rot_n4_g2}. Bottom panel: the corresponding structure constants. The vanishing of the structure constants indicates that this is an Abelian subgroup.}\label{fig:so4_2generators_easy}
\end{figure}

One of our found examples of a closed subalgebra with $N_g=2$ generators is shown in Figure~\ref{fig:so4_2generators_easy}. The top panels indicate that the two found generators can be roughly approximated as
\begin{subequations}
\begin{eqnarray}
\mathbb J_1 &\approx & 
\left(
\begin{array}{rrrr}
0 & 1 & 0 &  0 \\
-1 & 0 & 0 &  0 \\
0 & 0 & 0 &  0 \\
0 & 0 & 0 &  0
\end{array}
\right)
= {\mathbb O}_{12},
\\
\mathbb J_2 &\approx & 
\left(
\begin{array}{rrrr}
0 & 0 & 0 &  0 \\
0 & 0 & 0 &  0 \\
0 & 0 & 0 &  1 \\
0 & 0 & -1 &  0
\end{array}
\right)
= {\mathbb O}_{34},
\end{eqnarray}
\label{eq:SO4gen2}
\end{subequations}
in which we recognize the rotation generators ${\mathbb O}_{12}$ and ${\mathbb O}_{34}$ from (\ref{eq:groupgenerator}). As seen from the matrix forms in (\ref{eq:SO4gen2}), these two generators commute, since the two rotations are done in completely different planes. Therefore, the algebra formed by $\mathbb J_1$ and $\mathbb J_2$ is Abelian, which is independently verified by the bottom panel in Figure~\ref{fig:so4_2generators_easy}. 

\begin{figure}[t]
    \centering
    \includegraphics[width=0.47\textwidth]{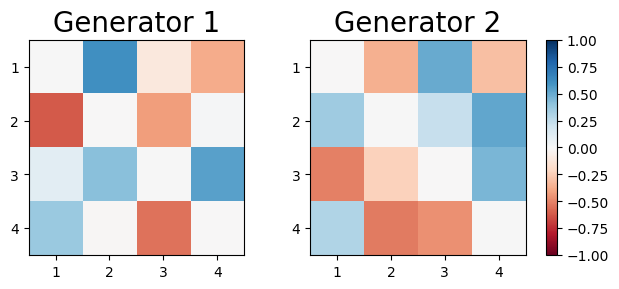} \\
    \includegraphics[width=0.3\textwidth]{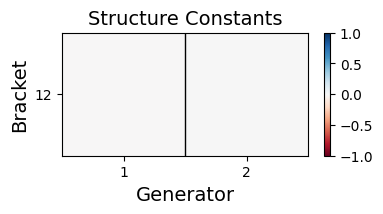}        
    \caption{The same as Figure~\ref{fig:so4_2generators_easy}, but for a different training run, still with $N_g=2$.}\label{fig:so4_2generators}
\end{figure}

One should keep in mind that we do not control the overall orientation of the generators found by our procedure. The example in Figure~\ref{fig:so4_2generators_easy} was judiciously chosen to be easily recognizable in terms of the canonical generators (\ref{eq:groupgenerator}). A generic training run typically returns the generator set in some random orientation, which, however, still preserves the commutation properties. One such generic example is shown in Figure~\ref{fig:so4_2generators}.
This time, the two found generators $\mathbb J_1$ and $\mathbb J_2$ are more general linear combinations of the six canonical generators ${\mathbb O}_{12}$, ${\mathbb O}_{13}$, ${\mathbb O}_{14}$, ${\mathbb O}_{23}$, ${\mathbb O}_{24}$, and ${\mathbb O}_{34}$ of the $O(4)$ group. Nevertheless, the found generators $\mathbb J_1$ and $\mathbb J_2$ still commute and form an Abelian two-dimensional subalgebra of the full symmetry group.

\subsection{Three generator subalgebras}
\label{sec:Rot_n4_g3}

\begin{figure}[t]
    \centering
    \includegraphics[width=0.47\textwidth]{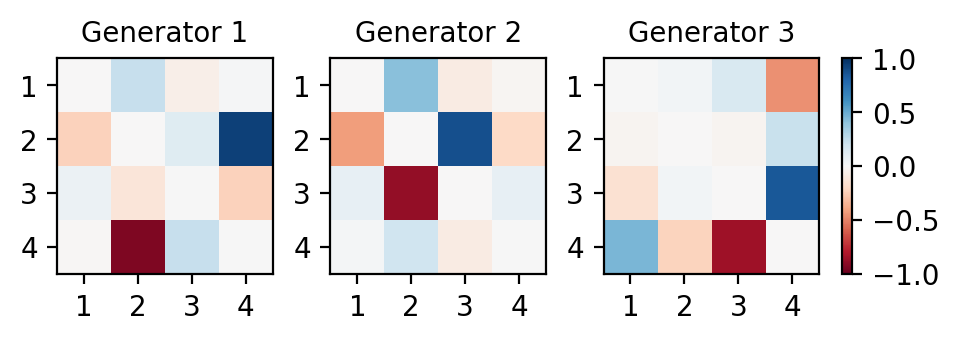}\\
    \includegraphics[width=0.3\textwidth]{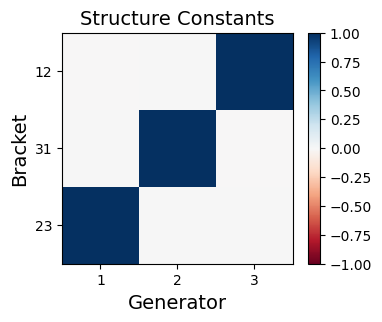}    
    \caption{Top row: the successfully learned generators for the $n=4$, $N_g=3$ exercise considered in Section \ref{sec:Rot_n4_g3}. Bottom panel: the corresponding structure constants, which can be identified as those of the SO(3) algebra (see text). }\label{fig:so4_3generators_easy}
\end{figure}

Next we discuss the discovered subalgebras with three generators ($N_g=3$). Since $SO(4)$ contains $SO(3)$ as a subgroup,\footnote{Since $SO(4) = SO(3) \otimes SO(3)$, the $SO(4)$ algebra is a  direct sum of two separate $SO(3)$ subalgebras.} we know that such subalgebras should exist, and indeed, we find such solutions, as seen in Figure~\ref{fig:lossn4}. As mentioned above, a generic training run produces the three generators in a random orientation. For ease of interpretation, in Figure~\ref{fig:so4_3generators_easy} we show a judiciously chosen example, in which the generators depicted in the top panels can be recognized to be approximately 
\beq
{\mathbb J}_{1} \approx {\mathbb O}_{24}, \qquad
{\mathbb J}_{2} \approx {\mathbb O}_{23},\qquad 
{\mathbb J}_{3} \approx {\mathbb O}_{34}.
\eeq
This algebra involves rotations primarily in the last three feature dimensions, while the first feature is largely unaffected by the symmetry. The bottom panel of Figure~\ref{fig:so4_3generators_easy} confirms that this is the $SO(3)$ algebra from eq.~(\ref{eq:SO3algebra}). Of course, a more generic training run results in a set of three generators that involve all four feature dimensions, but still have the same commutation relations.

\subsection{Four generator subalgebras}
\label{sec:Rot_n4_g4}

Figure~\ref{fig:lossn4} shows that our method finds an algebra with four generators as well. To see its origin theoretically, define a new generator basis in terms of sums and differences of pairs of commuting generators from the original basis (\ref{eq:groupgenerator}) \cite{hladik1999spinors}
\begin{subequations}
\begin{eqnarray}
{\mathbb S}_1 &\equiv \frac{1}{2}\left(
{\mathbb O}_{34}+{\mathbb O}_{12}\right),
\quad
{\mathbb D}_1 &\equiv \frac{1}{2}\left(
{\mathbb O}_{34}-{\mathbb O}_{12}\right), 
\\
{\mathbb S}_2 &\equiv \frac{1}{2}\left(
{\mathbb O}_{42}+{\mathbb O}_{13}\right),
\quad
{\mathbb D}_2 &\equiv \frac{1}{2}\left(
{\mathbb O}_{42}-{\mathbb O}_{13}\right), 
\\
{\mathbb S}_3 &\equiv \frac{1}{2}\left(
{\mathbb O}_{23}+{\mathbb O}_{14}\right),
\quad
{\mathbb D}_3 &\equiv \frac{1}{2}\left(
{\mathbb O}_{23}-{\mathbb O}_{14}\right).
\end{eqnarray}
\end{subequations}
This change of basis decouples the algebra (\ref{eq:rotcommutator}) of the original generators as follows
\begin{subequations}
\begin{eqnarray}
\left[ {\mathbb S}_i, {\mathbb S}_j\right] &=& -\,\epsilon_{ijk}\,{\mathbb S}_k, \label{algebraofSs}\\
\left[ {\mathbb D}_i, {\mathbb D}_j\right] &=& -\,\epsilon_{ijk}\,{\mathbb D}_k, \label{algebraofDs}\\
\left[ {\mathbb S}_i, {\mathbb D}_j\right] &=& 0.
\label{algebraofSsAndDs}
\end{eqnarray}
\end{subequations}
\begin{figure}[t]
    \centering 
    \includegraphics[width=0.47\textwidth]{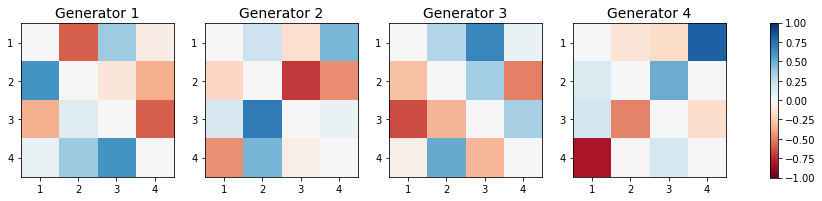}\\
    \includegraphics[width=0.3\textwidth]{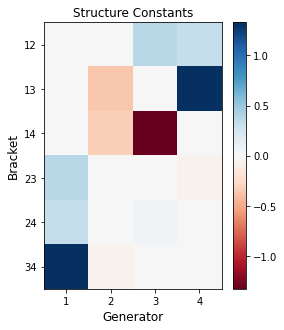}
    \caption{Top row: the successfully learned generators for the $n=4$, $N_g=4$ exercise considered in Section \ref{sec:Rot_n4_g4}. Bottom panel: the corresponding structure constants.  }\label{fig:so4_4generators}
\end{figure}
Therefore, a closed algebra of four generators can be formed either from the set of three $\mathbb S$'s plus any one of the $\mathbb D$'s, or from the set of three $\mathbb D$'s plus any one of the $\mathbb S$'s. In either case, three of the generators will satisfy $SO(3)$-type commutation relations, while the fourth one will commute with everyone else. This expectation is confirmed in Figure~\ref{fig:so4_4generators}, which shows our usual representation of the found generators and their algebra for one representative result from a training run with $N_g=4$. We observe that the found generators are approximately
\begin{subequations}
\begin{eqnarray}
{\mathbb J}_1 &\equiv& \frac{1}{2}\left(
{\mathbb O}_{21}+{\mathbb O}_{43}\right) = -\, {\mathbb S}_1,
\\
{\mathbb J}_2 &\equiv& \frac{1}{2}\left(
{\mathbb O}_{32}+{\mathbb O}_{14}\right) = -\, {\mathbb D}_3, 
\\
{\mathbb J}_3 &\equiv& \frac{1}{2}\left(
{\mathbb O}_{13}+{\mathbb O}_{42}\right)= \, {\mathbb S}_2, 
\\
{\mathbb J}_4 &\equiv& \frac{1}{2}\left(
{\mathbb O}_{14}+{\mathbb O}_{23}\right)= \, {\mathbb S}_3.
\end{eqnarray}
\end{subequations}
Therefore, ${\mathbb J}_1$, ${\mathbb J}_3$ and ${\mathbb J}_4$ form an SO(3) algebra as implied by eq.~(\ref{algebraofSs}), and furthermore, all three of them commute with ${\mathbb J}_2$, as implied by eq.~(\ref{algebraofSsAndDs}). This pattern is precisely what we observe in the lower panel of Figure~\ref{fig:so4_4generators}.

\subsection{Six generator algebras}
\label{sec:Rot_n4_g6}

\begin{figure}[t]
    \centering
    \includegraphics[width=0.48\textwidth]{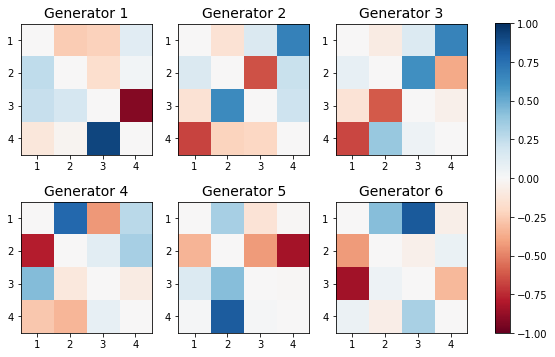}\\
    \includegraphics[width=0.3\textwidth]{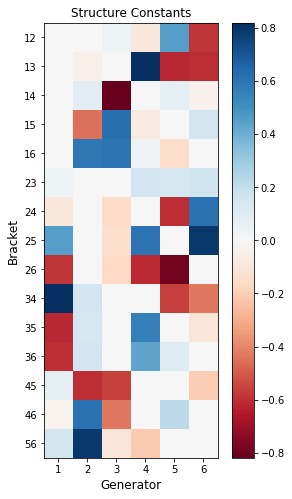} 
    \caption{Top row: the successfully learned generators for the $n=4$, $N_g=6$ exercise considered in Section \ref{sec:Rot_n4_g6}. Bottom panel: the corresponding structure constants.  }\label{fig:so4_6generators}
\end{figure}

Finally, we get to the case of $N_g$ which will reveal the full algebra of $SO(4)$. Figure~\ref{fig:so4_6generators} shows the result from a representative training run seeking a closed algebra with $N_g=6$ generators. Among the set of learned generators we can approximately recognize 
${\mathbb J}_1 \approx {\mathbb O}_{43}$,
${\mathbb J}_2 \approx -{\mathbb D}_{3}$,
${\mathbb J}_3 \approx {\mathbb S}_{3}$,
${\mathbb J}_4 \approx {\mathbb O}_{12}$,
${\mathbb J}_5 \approx {\mathbb O}_{42}$ and
${\mathbb J}_6 \approx {\mathbb O}_{13}$.

The analysis of the last three sections can be readily extended to even higher dimensions ($n\ge 5$). We have checked a few more values of $n$ and the method works each time --- we obtain valid finite symmetry transformations which preserve the oracle (\ref{eq:length}), we find the closed algebra of the full set of $n(n-1)/2$ generators of the orthogonal $O(n)$ group, as well as valid subalgebras. For fun, in Figure~\ref{fig:SO10} we depict our derived 45 generators of the $SO(10)$ group.

\begin{figure}[t]
    \centering
    \includegraphics[width=0.47\textwidth]{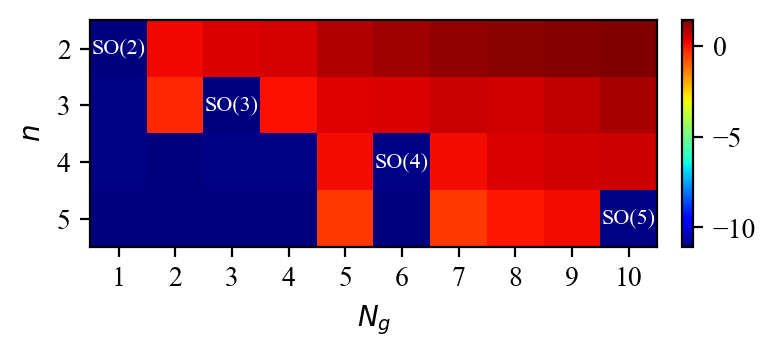}
    \caption{Results from our search for Lie algebras of transformations preserving the Euclidean oracle (\ref{eq:length}) in $n=2,3,4,5$ dimensions and for different number of generators $N_g$. The cells are color coded by the base-10 logarithm of the lowest value of the loss attained during training.  }\label{fig:subalgebras}
\end{figure}

In conclusion of our discussion of orthogonal groups, in Figure~\ref{fig:subalgebras} we summarize our results for the found closed algebras and subalgebras for $n\le 5$ and different number of generators $N_g$. Each cell in the table represents a separate exercise at a fixed number of dimensions $n$ and for a fixed number of generators $N_g$. The cells are color coded by the base-10 logarithm of the lowest value of the loss attained during training. The training is terminated once the loss reaches $10^{-12}$, therefore blue cells correspond to successful training runs resulting in closed algebras. The right-most blue cell in each row corresponds to the full algebra, in this case $SO(n)$, as indicated. The preceding blue cells correspond to various subalgebras. In fact we did not anticipate the existence of the $N_g=4$ subalgebras in the case of $n=4$ and $n=5$, but our model surprised us. 

\section{Lorentz Transformations in Four Dimensional Minkowski Space}
\label{sec:Lorentz}

In this section we consider the four-dimensional Minkowski spacetime $(t,x,y,z)$ (in natural units with $c=1$). The usual Lorentz transformations preserve the quadratic form
\beq
\varphi(t,x,y,z) = t^2-x^2-y^2-z^2, \quad (t,x,y,z)\in {\mathbb R}^4,
\label{eq:oracleLor}
\eeq
hence this will be our oracle in this section. The four input features are
\beq
x^{(0)}=t, \quad
x^{(1)}=x, \quad
x^{(2)}=y, \quad
x^{(3)}=z,
\eeq
where in keeping with the standard physics notation we start labelling the features from 0.

Our analysis proceeds as usual. First, we find symmetry transformations which are in general combinations of boosts and rotations. Upon inspection, we verify that they have the desired symmetry properties
\beq
{\mathbb F}_{0i}={\mathbb F}_{i0},~~
{\mathbb F}_{ij}=-{\mathbb F}_{ji},~~
{\mathbb F}_{00}={\mathbb F}_{ii}=0,~~
\forall i=1,2,3.
\nonumber
\eeq

Next we analyze the algebras of generators. Before presenting the numerical results, for the reader's convenience we summarize some relevant information about the mathematical structure of the Lorentz group $O(1,3)$. It has six generators: the three generators of boosts ${\mathbb K}_i$,
\begin{subequations}
\begin{eqnarray}
{\mathbb K}_1 &=& 
\left(
\begin{array}{rrrr}
    0 & 1 & 0 & 0 \\ 
    1 & 0 & 0 & 0 \\ 
    0 & 0 & 0 & 0 \\ 
    0 & 0 & 0 & 0
\end{array}     
\right), \\
{\mathbb K}_2 &=& 
\left(
\begin{array}{rrrr}
    0 & 0 & 1 & 0 \\ 
    0 & 0 & 0 & 0 \\ 
    1 & 0 & 0 & 0 \\ 
    0 & 0 & 0 & 0
\end{array}     
\right), \\
{\mathbb K}_3 &=& 
\left(
\begin{array}{rrrr}
    0 & 0 & 0 & 1 \\ 
    0 & 0 & 0 & 0 \\ 
    0 & 0 & 0 & 0 \\ 
    1 & 0 & 0 & 0
\end{array}     
\right),
\end{eqnarray}
\end{subequations}
and the three generators of rotations, ${\mathbb L}_i$, given by
\begin{subequations}
\begin{eqnarray}
{\mathbb L}_1 &=& 
\left(
\begin{array}{rrrr}
    0 & 0 & 0 & 0 \\ 
    0 & 0 & 0 & 0 \\ 
    0 & 0 & 0 & -1 \\ 
    0 & 0 & 1 & 0
\end{array}     
\right),\\
{\mathbb L}_2 &=& 
\left(
\begin{array}{rrrr}
    0 & 0 & 0 & 0 \\ 
    0 & 0 & 0 & 1 \\ 
    0 & 0 & 0 & 0 \\ 
    0 & -1 & 0 & 0
\end{array}     
\right), \\
{\mathbb L}_3 &=& 
\left(
\begin{array}{rrrr}
    0 & 0 & 0 & 0 \\ 
    0 & 0 & -1 & 0 \\ 
    0 & 1 & 0 & 0 \\ 
    0 & 0 & 0 & 0
\end{array}     
\right),
\end{eqnarray}
\end{subequations}
The full Lorentz algebra can be summarized as 
\begin{eqnarray}
[{\mathbb L}_i, {\mathbb L}_j]  &=&  \epsilon_{ijk}\,{\mathbb L}_k, \\ [2mm]
[{\mathbb L}_i, {\mathbb K}_j]  &=&  \epsilon_{ijk}\, {\mathbb K}_k, \\ [2mm]
[{\mathbb K}_i, {\mathbb K}_j]  &=&  - \epsilon_{ijk}\, {\mathbb L}_k. 
\end{eqnarray}
The subgroup structure is most easily seen in a different basis,
\begin{subequations}
\begin{eqnarray}
{\mathbb X}_1 &=& {\mathbb K}_1 - {\mathbb L}_2, \label{eq:X1def}\\ [2mm]
{\mathbb X}_2 &=& {\mathbb K}_2 + {\mathbb L}_1, \label{eq:X2def}\\ [2mm]
{\mathbb X}_3 &=& {\mathbb K}_3, \\ [2mm]
{\mathbb X}_4 &=& {\mathbb L}_3, \\ [2mm]
{\mathbb X}_5 &=& {\mathbb L}_2, \\ [2mm]
{\mathbb X}_6 &=& {\mathbb L}_1.
\end{eqnarray}
\label{eq:algebraX}
\end{subequations}

\begin{table}[t!]
\centering
\scalebox{1.0}{
\renewcommand\arraystretch{1.8}
\begin{tabular}{||c||c|c|c|c|c|c||}
\cline{2-7}
 \multicolumn{1}{c||}{}& ${\mathbb K}_1-{\mathbb L}_2$ & ${\mathbb K}_2+{\mathbb L}_1$ & ${\mathbb K}_3$ & ${\mathbb L}_3$ & ${\mathbb L}_2$ & ${\mathbb L}_1$ \\ \cline{2-7}
\multicolumn{1}{c||}{} & ${\mathbb X}_1$ & ${\mathbb X}_2$ & ${\mathbb X}_3$ & ${\mathbb X}_4$ & ${\mathbb X}_5$ & ${\mathbb X}_6$ \\ 
\hline
\hline
~${\mathbb X}_1$~ &    0   &    0   &  $-{\mathbb X}_1$  & $-{\mathbb X}_2$ & $+{\mathbb X}_3$ & $+{\mathbb X}_4$\\ \hline
${\mathbb X}_2$&    0   &    0   &  $-{\mathbb X}_2$  & $+{\mathbb X}_1$ & $+{\mathbb X}_4$ & $-{\mathbb X}_3$\\ \hline
${\mathbb X}_3$& $+{\mathbb X}_1$ & $+{\mathbb X}_2$ &    0     &   0    & $-{\mathbb X}_1-{\mathbb X}_5$ & ${\mathbb X}_2-{\mathbb X}_6$ \\ \hline
${\mathbb X}_4$& $+{\mathbb X}_2$ & $-{\mathbb X}_1$ &    0     &   0    & $-{\mathbb X}_6$ & $+{\mathbb X}_5$ \\ \hline
${\mathbb X}_5$& $-{\mathbb X}_3$ & $-{\mathbb X}_4$ &$+{\mathbb X}_1+{\mathbb X}_5$& $+{\mathbb X}_6$ &   0    & $-{\mathbb X}_4$\\ \hline
${\mathbb X}_6$& $-{\mathbb X}_4$ & $+{\mathbb X}_3$ &$-{\mathbb X}_2+{\mathbb X}_6$& $-{\mathbb X}_5$ & $+{\mathbb X}_4$ &  0 \\ 
\hline\hline
\end{tabular}
}
\caption{The Lorentz algebra in terms of the generators (\ref{eq:algebraX}). The cells show the result from commuting one of the generators in the leftmost column with one of the generators in the top row.  
\label{tab:LorentzAlgebraX}}
\end{table}

The resulting algebra in the ${\mathbb X}_i$ basis is summarized in Table~\ref{tab:LorentzAlgebraX}. By inspection, we see that the Lorentz algebra has several non-trivial subalgebras.

\subsection{Two generator subalgebras} 
\label{sec:Lorentz_2gen}

There are several two-dimensional subalgebras:
\begin{itemize}
\item {\em Abelian subalgebras.} There are two abelian subalgebras: the first one is generated by the generator set $\{{\mathbb X}_1,{\mathbb X}_2\}$,  
while the second is generated by $\{{\mathbb X}_3,{\mathbb X}_4\}$.
A representative example of the former case is shown in Figure~\ref{fig:Lor_2generators}. We recognize the two found generators to be approximately 
\begin{subequations}
\begin{eqnarray}
{\mathbb J}_1 &\approx& -\left( {\mathbb K}_3 -  {\mathbb L}_1 \right), \\ [2mm]
{\mathbb J}_2 &\approx& -\left( {\mathbb K}_1 +  {\mathbb L}_3 \right).
\end{eqnarray}
\end{subequations}
This result is in agreement with eqs.~(\ref{eq:X1def}) and (\ref{eq:X2def}) --- the transformations are combinations of boosts along and rotations about two of the axes, in this case $x$ and $z$.
\begin{figure}[t]
    \centering
    \includegraphics[width=0.45\textwidth]{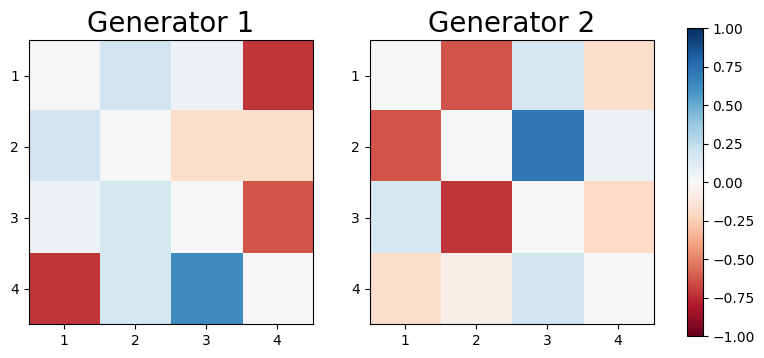}
\\
    \includegraphics[width=0.3\textwidth]{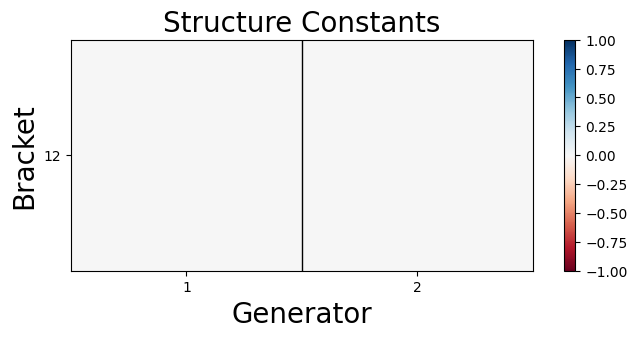}    
    \caption{Top row: successfully learned generators for the $n=4$, $N_g=2$ exercise considered in Section \ref{sec:Lorentz_2gen}, using the pseudo-Euclidean oracle  (\ref{eq:oracleLor}). Bottom panel: the corresponding structure constants. This example shows an abelian subalgebra. }\label{fig:Lor_2generators}
\end{figure}
\begin{figure}[t]
    \centering 
    \includegraphics[width=0.45\textwidth]{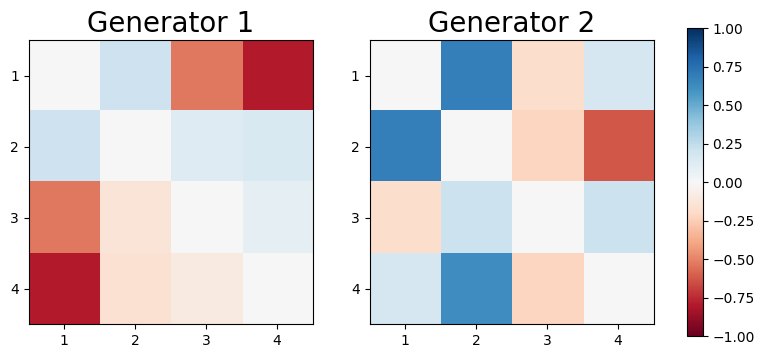}\\
    \includegraphics[width=0.3\textwidth]{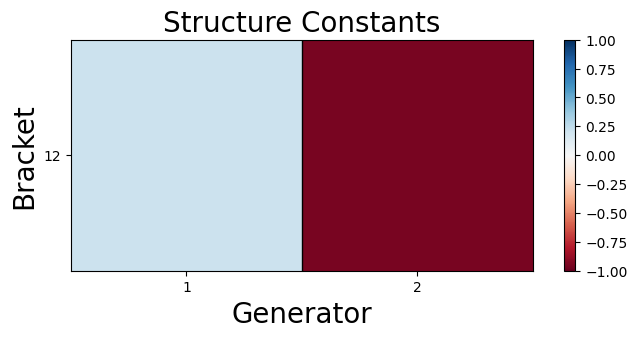}        
    \caption{The same as Fig.~\ref{fig:Lor_2generators}, but showing a non-abelian subalgebra found by our method. }\label{fig:Lor_2generators_nonabelian}
\end{figure}
\item {\em Nonabelian subalgebras.} As seen in Table~\ref{tab:LorentzAlgebraX}, the Lie algebra of the Lorentz group has two nonabelian subalgebras, generated by $\{{\mathbb X}_1, {\mathbb X}_3\}$ and $\{{\mathbb X}_2,{\mathbb X}_3\}$, respectively. A representative training example for this nonabelian case is shown in Figure~\ref{fig:Lor_2generators_nonabelian}. The top panels show that the two found generators are approximately 
\begin{subequations}
\begin{eqnarray}
{\mathbb J}_1 &\approx& -\, {\mathbb K}_3 = -\, {\mathbb X}_3, \\ [2mm]
{\mathbb J}_2 &\approx& {\mathbb K}_1 -  {\mathbb L}_2 = {\mathbb X}_1,
\end{eqnarray}
\end{subequations}
\end{itemize}
while the bottom panel confirms that their Lie bracket is approximately given by $[{\mathbb J}_1, {\mathbb J}_2]\approx 
[-{\mathbb X}_3,{\mathbb X}_1]=-{\mathbb X}_1=-\, {\mathbb J}_2$, as expected from Table~\ref{tab:LorentzAlgebraX}.

\subsection{Three generator subalgebras}
\label{sec:Lorentz_3gen}

Table~\ref{tab:LorentzAlgebraX} reveals several three-dimensional subalgebras:
\begin{itemize}
\item {\em SO(3).}  The set $\{{\mathbb X}_4,{\mathbb X}_5,{\mathbb X}_6\}$ generates a subalgebra isomorphic to the Lie algebra of the $SO(3)$ group of rotations in three dimensions, see Section~\ref{sec:rotations3dim}.
A representative training example is shown in Figure~\ref{fig:Lor_3generators_so3}.
\begin{figure}[t]
    \centering
    \includegraphics[width=0.48\textwidth]{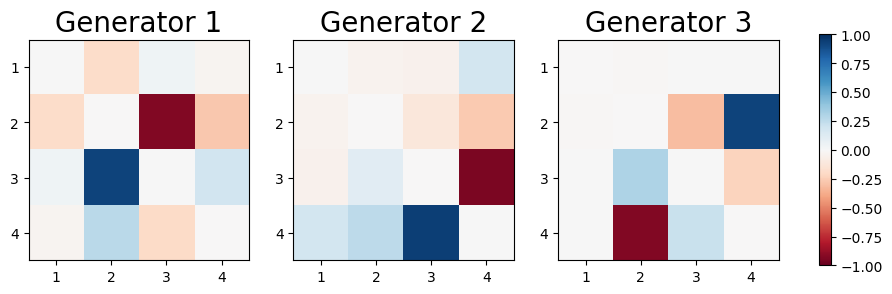} \\
    \includegraphics[width=0.3\textwidth]{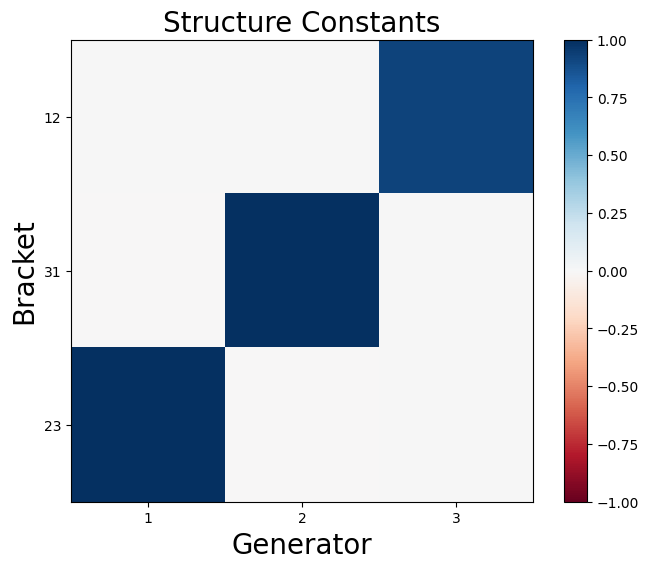}        
    \caption{
    Top row: the successfully learned generators for the $n=4$, $N_g=3$ exercise considered in Section \ref{sec:Lorentz_3gen}, using the pseudo-Euclidean oracle  (\ref{eq:oracleLor}). Bottom panel: the corresponding structure constants, which can be identified as those of an {\em SO(3)} type subalgebra. }\label{fig:Lor_3generators_so3}
\end{figure}
The generators can be recognized as
\begin{subequations}
\begin{eqnarray}
{\mathbb J}_1 &\approx& {\mathbb L}_3, \\ [2mm]
{\mathbb J}_2 &\approx& {\mathbb L}_1, \\ [2mm]
{\mathbb J}_3 &\approx& {\mathbb L}_2.
\end{eqnarray}
\end{subequations}
In addition, we also found examples with subalgebras isomorphic to $so(3)$ consisting of one rotation and two boosts, or one boost and two rotations. 
\item {\em Hom(2).} The set $\{{\mathbb X}_1,{\mathbb X}_2,{\mathbb X}_3\}$ generates a subalgebra isomorphic to the Lie algebra of Hom(2), the group of Euclidean homotheities:
\begin{equation}
[{\mathbb X}_1, {\mathbb X}_2] = 0,
~
[{\mathbb X}_3, {\mathbb X}_1] = {\mathbb X}_1,
~
[{\mathbb X}_2, {\mathbb X}_3] = -{\mathbb X}_2.
\label{eq:homalgebra}
\end{equation}
A representative training example is shown in Figure~\ref{fig:Lor_3generators_hom}. From the panels in the top row we recognize the found three generators as
\begin{subequations}
\begin{eqnarray}
{\mathbb J}_1 &\approx& +\, {\mathbb K}_2 -\, {\mathbb L}_1, \\ [2mm]
{\mathbb J}_2 &\approx& -\, {\mathbb K}_1 -\, {\mathbb L}_2, \\ [2mm]
{\mathbb J}_3 &\approx& +\, {\mathbb K}_3.
\end{eqnarray}
\end{subequations}
The bottom panel in Figure~\ref{fig:Lor_3generators_hom} confirms that their algebra is approximately that of eq.~(\ref{eq:homalgebra}).
\begin{figure}[t]
    \centering
    \includegraphics[width=0.48\textwidth]{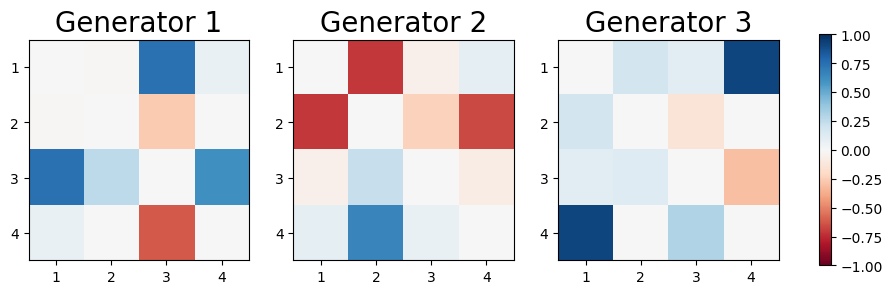} \\
    \includegraphics[width=0.3\textwidth]{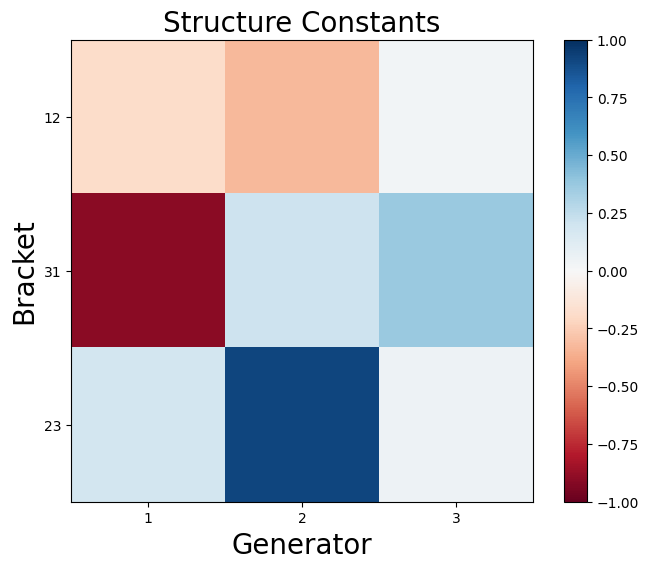}    
    \caption{The same as Fig.~\ref{fig:Lor_3generators_so3}, but showing an example of a {\em Hom(2)} subalgebra (\ref{eq:homalgebra}). }\label{fig:Lor_3generators_hom}
\end{figure}

\begin{figure}[t]
    \centering
    \includegraphics[width=0.48\textwidth]{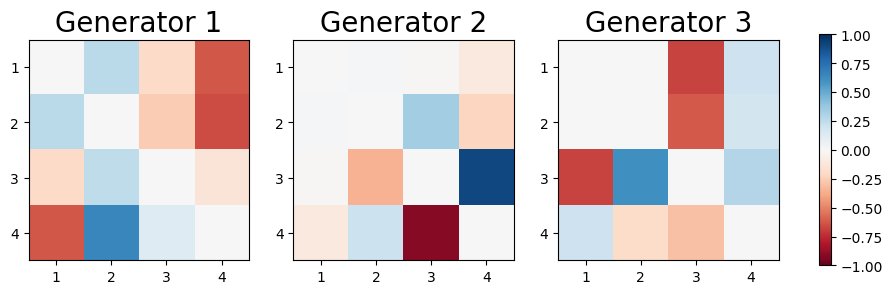}\\
    \includegraphics[width=0.3\textwidth]{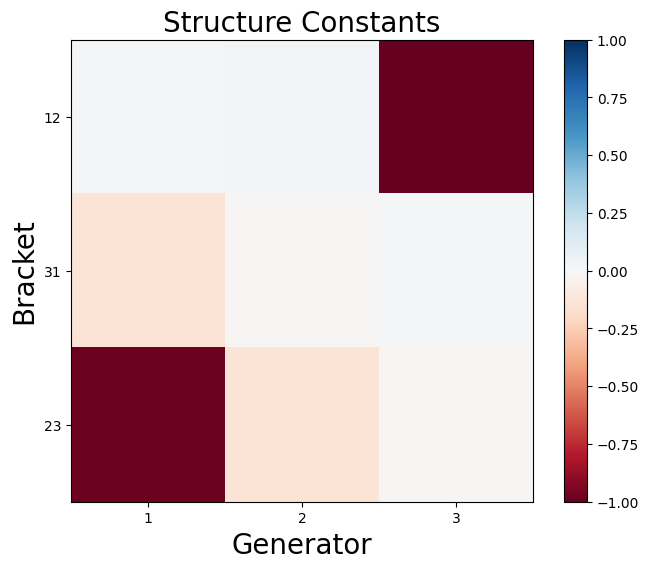}    
    \caption{The same as Fig.~\ref{fig:Lor_3generators_so3}, but showing an example of a {\em E(2)} subalgebra (\ref{eq:e2algebra}).  }\label{fig:Lor_3generators_e2}
\end{figure}
\item {\em E(2).} The set $\{{\mathbb X}_1,{\mathbb X}_2,{\mathbb X}_4\}$ generates a subalgebra isomorphic to the Lie algebra of $E(2)$, the Euclidean group:
\begin{equation}
[{\mathbb X}_1, {\mathbb X}_2] = 0,
~
[{\mathbb X}_4, {\mathbb X}_1] = {\mathbb X}_2,
~
[{\mathbb X}_2, {\mathbb X}_4] = {\mathbb X}_1.
\label{eq:e2algebra}
\end{equation}
A representative example is shown in Figure~\ref{fig:Lor_3generators_e2}, from which we recognize the found three generators as
\begin{subequations}
\begin{eqnarray}
{\mathbb J}_1 &\approx& -\, {\mathbb K}_3 -\, {\mathbb L}_2, \\ [2mm]
{\mathbb J}_2 &\approx& -\, {\mathbb L}_1, \\ [2mm]
{\mathbb J}_3 &\approx& -\, {\mathbb K}_2 +\, {\mathbb L}_3.
\end{eqnarray}
\end{subequations}
As expected, two of the generators involve combinations of boosts along and rotations about two of the axes, in this case $y$ and $z$, while the third generator is a rotation about the remaining axis, namely $x$. The bottom panel in Figure~\ref{fig:Lor_3generators_e2} shows that the resulting algebra is approximately that of Eq.~(\ref{eq:e2algebra}).
\item {\em Bianchi VII${}_a$.} The set $\{{\mathbb X}_1,{\mathbb X}_2,{\mathbb X}_3+a{\mathbb X}_4\}$ with $a\ne 0$ generates a Bianchi VII${}_a$ subalgebra. A representative training example is shown in Figure~\ref{fig:Lor_3generators_bia}, from which we recognize the found three generators as
\begin{subequations}
\begin{eqnarray}
{\mathbb J}_1 &\approx& -\, {\mathbb K}_3 -\, {\mathbb L}_3 = -({\mathbb X}_3+{\mathbb X}_4), \\ [2mm]
{\mathbb J}_2 &\approx& +\, {\mathbb K}_1 -\, {\mathbb L}_2 = {\mathbb X}_1, \\ [2mm]
{\mathbb J}_3 &\approx& -\, {\mathbb K}_2 -\, {\mathbb L}_1 = -\, {\mathbb X}_2.
\end{eqnarray}
\end{subequations}
This set corresponds to a Bianchi VII${}_a$ subalgebra with $a=1$, whose structure constants are indeed consistent with the lower panel in Figure~\ref{fig:Lor_3generators_bia}.
\begin{figure}[t]
    \centering
    \includegraphics[width=0.48\textwidth]{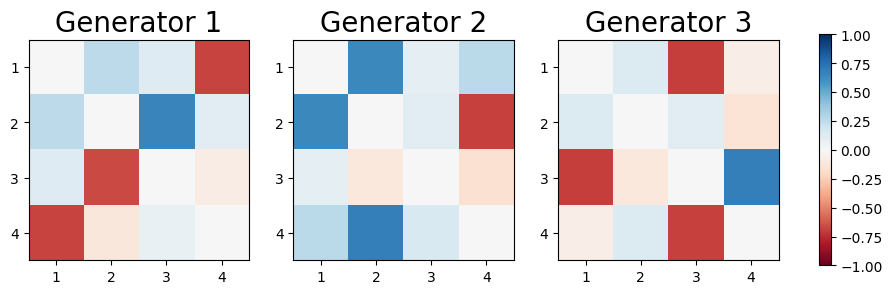}\\
    \includegraphics[width=0.3\textwidth]{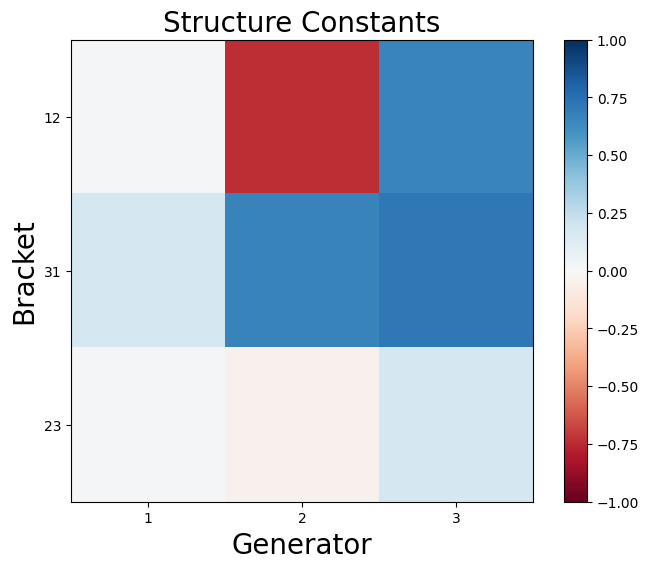}        \caption{
    The same as Fig.~\ref{fig:Lor_3generators_so3}, but showing an example of a {\em Bianchi VII${}_a$} subalgebra. }\label{fig:Lor_3generators_bia}
\end{figure}
\item {\em SL(2,R).} The set $\{{\mathbb X}_1,{\mathbb X}_3,{\mathbb X}_5\}$ generates a subalgebra isomorphic to the Lie algebra of $SL(2,R)$, the group of isometries of the hyperbolic plane:
\begin{subequations}
\begin{eqnarray}
 [{\mathbb X}_1, {\mathbb X}_3] &=& -\, {\mathbb X}_1, \\ [2mm]
 [{\mathbb X}_5, {\mathbb X}_1] &=& -\, {\mathbb X}_3, \\ [2mm]
 [{\mathbb X}_3, {\mathbb X}_5] &=& -\, {\mathbb X}_1  -\, {\mathbb X}_5.
\end{eqnarray}
\label{eq:SL2Ralgebra}
\end{subequations}

A representative training example is shown in Figure~\ref{fig:Lor_3generators_sl2r}, from which we recognize the found three generators as
\begin{subequations}
\begin{eqnarray}
{\mathbb J}_1 &\approx& {\mathbb K}_1 - {\mathbb L}_3, \\ [2mm]
{\mathbb J}_2 &\approx& -\, {\mathbb L}_3, \\ [2mm]
{\mathbb J}_3 &\approx& {\mathbb K}_2.
\end{eqnarray}
\end{subequations}
\begin{figure}[t]
    \centering
    \includegraphics[width=0.48\textwidth]{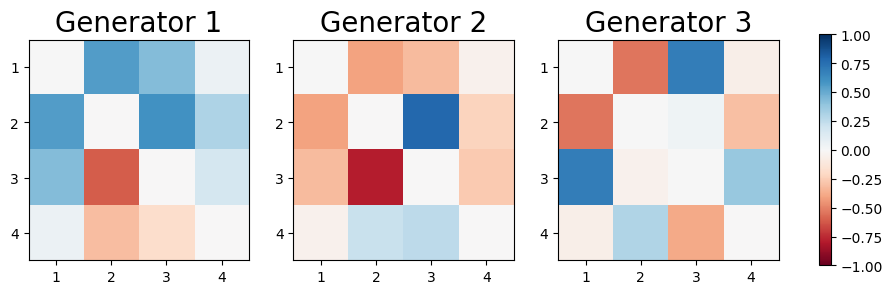}\\
    \includegraphics[width=0.3\textwidth]{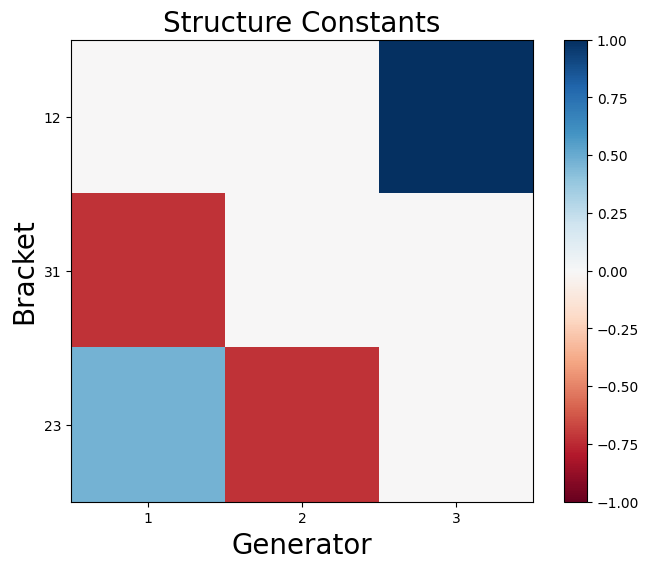}    
    \caption{
    The same as Fig.~\ref{fig:Lor_3generators_so3}, but showing an example of an {\em SL(2,R)} subalgebra (\ref{eq:SL2Ralgebra}). }\label{fig:Lor_3generators_sl2r}
\end{figure}
It is easy to verify that their structure constants given in the lower panel of Figure~\ref{fig:Lor_3generators_sl2r} are consistent with Eq.~(\ref{eq:SL2Ralgebra}).
\end{itemize}

\subsection{Four generator subalgebras}
\label{sec:Lorentz_4gen}

\begin{figure}[t]
    \centering
    \includegraphics[width=0.48\textwidth]{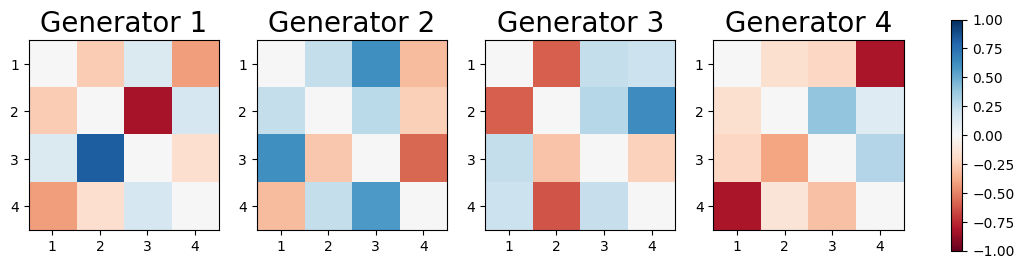}\\
    \includegraphics[width=0.25\textwidth]{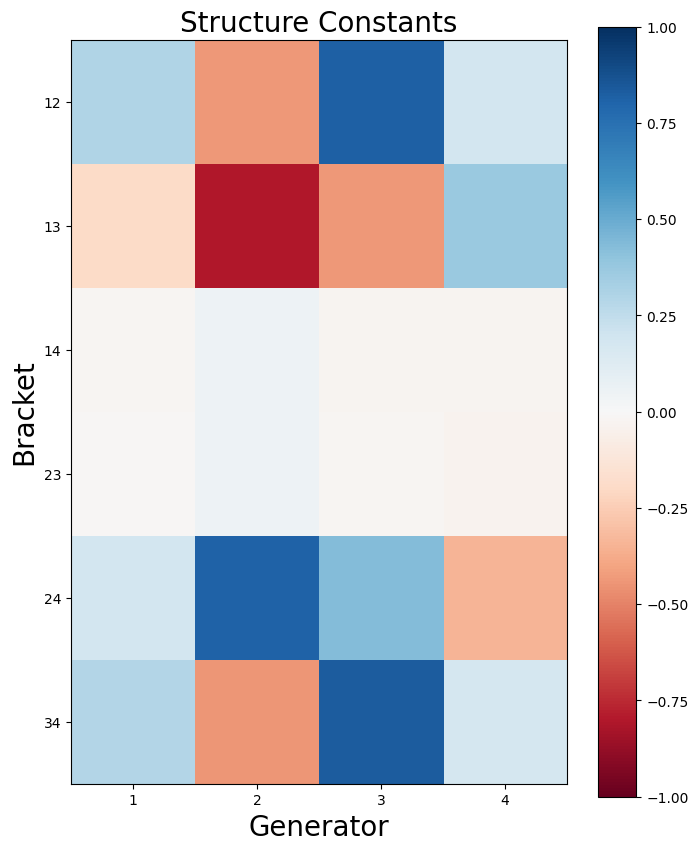}    
    \caption{
    Top row: the successfully learned generators for the $n=4$, $N_g=4$ exercise considered in Section \ref{sec:Lorentz_4gen}, using the pseudo-Euclidean oracle  (\ref{eq:oracleLor}). Bottom panel: the corresponding structure constants.}\label{fig:Lor_4generators}
\end{figure}

The only four-dimensional subalgebra is generated by $\{{\mathbb X}_1,{\mathbb X}_2,{\mathbb X}_3,{\mathbb X}_4\}$. A representative example is shown in Figure~\ref{fig:Lor_4generators}. The four found generators are given by 
\begin{subequations}
\begin{eqnarray}
{\mathbb J}_1 &\approx& {\mathbb L}_3 = {\mathbb X}_4, \\ [2mm]
{\mathbb J}_2 &\approx& {\mathbb K}_2 + {\mathbb L}_1 = {\mathbb X}_2, \\ [2mm]
{\mathbb J}_3 &\approx& -\, {\mathbb K}_1 + {\mathbb L}_2 = -\, {\mathbb X}_1, \\ [2mm]
{\mathbb J}_4 &\approx& -\, {\mathbb K}_3 = -\, {\mathbb X}_3.
\end{eqnarray}
\end{subequations}
The lower panel of Figure~\ref{fig:Lor_4generators} illustrates the corresponding structure constants. The results are consistent with Table~\ref{tab:LorentzAlgebraX}. For example, ${\mathbb J}_1$ and ${\mathbb J}_4$ commute because $[{\mathbb X}_3,{\mathbb X}_4]=[{\mathbb K}_3,{\mathbb L}_3]=0$. Similarly, ${\mathbb J}_2$ and ${\mathbb J}_3$ commute due to $[{\mathbb X}_1,{\mathbb X}_2]=0$. The results shown in the remaining rows in the lower panel of Figure~\ref{fig:Lor_4generators} can be analogously verified with the help of Table~\ref{tab:LorentzAlgebraX}.

\subsection{Six generator algebras}
\label{sec:Lorentz_6gen}

\begin{figure}[t]
    \centering
    \includegraphics[width=0.45\textwidth]{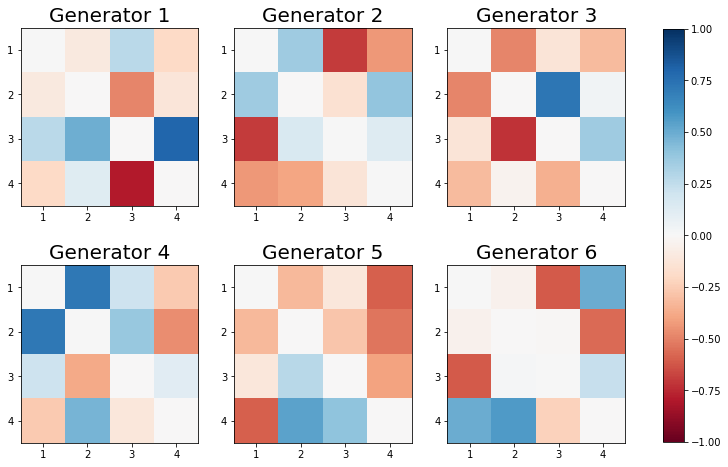} \\
    \includegraphics[width=0.3\textwidth]{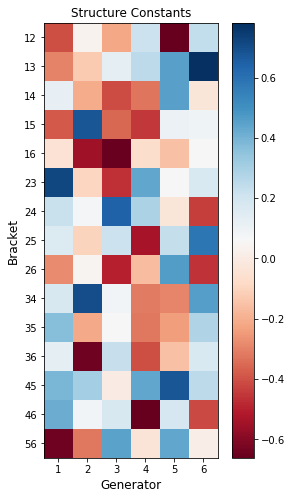}
    \caption{Top row: the successfully learned generators for the $n=4$, $N_g=6$ exercise considered in Section \ref{sec:Lorentz_6gen}, using the pseudo-Euclidean oracle (\ref{eq:oracleLor}). Bottom panel: the corresponding structure constants. } \label{fig:Lor_6generators}
\end{figure}

Our method is capable of finding the full six-dimensional algebra as well. The result is shown in Figure~\ref{fig:Lor_6generators}, and can be roughly identified as
\begin{subequations}
\begin{eqnarray}
{\mathbb J}_1 &\approx -\, {\mathbb L}_1, \quad 
{\mathbb J}_2 \approx -\, {\mathbb K}_2, \quad 
{\mathbb J}_3 \approx -\, {\mathbb L}_3, \\ [2mm]
{\mathbb J}_4 &\approx +\, {\mathbb K}_1, \quad
{\mathbb J}_5 \approx -\, {\mathbb K}_3, \quad 
{\mathbb J}_6 \approx -\, {\mathbb L}_2. 
\end{eqnarray}
\end{subequations}
The corresponding structure constants are shown in the lower panel of Figure~\ref{fig:Lor_6generators}.

\section{Squeeze Mapping in Two Dimensions}
\label{sec:squeeze}

\begin{figure}[t]
    \centering
    \includegraphics[width=0.45\textwidth]{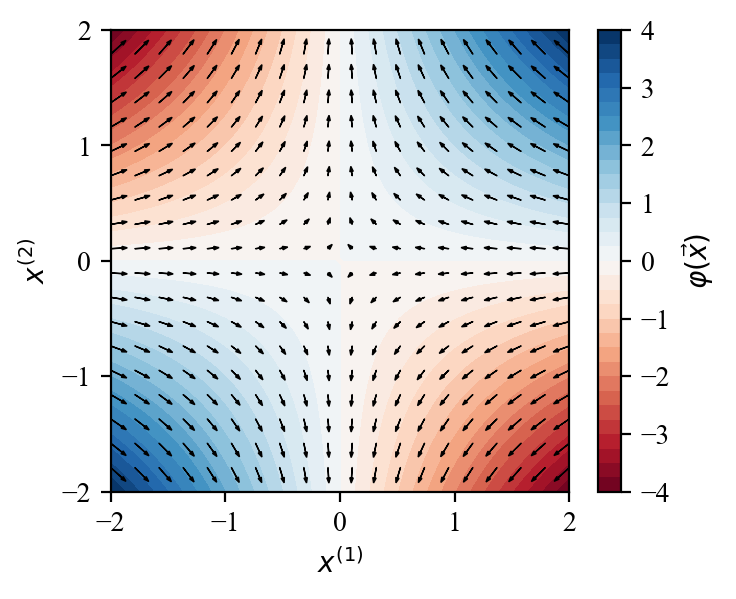}
    \caption{Color heatmap showing the values of the oracle (\ref{eq:oracleXY}) in the $(x^{(1)}, x^{(2)})$ Cartesian plane. The superimposed vector field represents the symmetry found by our method. }\label{fig:XYsymmetry}
\end{figure}

Consider again two dimensions, but now let the oracle return the product of the two input features:
\beq
\varphi(\mathbf x) = x^{(1)}\, x^{(2)}.
\label{eq:oracleXY}
\eeq
This oracle function is illustrated in Figure~\ref{fig:XYsymmetry} as a color heatmap representing the oracle values in the $(x^{(1)}, x^{(2)})$ Cartesian plane. We see that the set of points with the same oracle values form hyperbolas. 

Proceeding as before, our method finds the symmetry transformation illustrated with the vector field in Figure~\ref{fig:XYsymmetry}.
The corresponding generator is
\beq
\mathbb{J} = \left(
\begin{array}{rr}
-1 & 0 \\
0 & 1
\end{array}
\right).
\label{eq:XYgenerator}
\eeq
We verified that in this example the method is unable to find more than one generator.

These results are precisely what one would expect. The symmetry which preserves the oracle (\ref{eq:oracleXY}) is the squeeze mapping
\beq
\mathbb F = 
\left(
\begin{array}{cc}
\frac{1}{\ell} & 0 \\
0          & \ell
\end{array}
\right),
\eeq
where $\ell$ is a scale factor. Considering infinitesimal transformations (\ref{eq:rotinf2}) with $\ell = 1 + \varepsilon$ immediately leads to the generator (\ref{eq:XYgenerator}).

\section{Discontinuous Oracles}
\label{sec:discontinuous}

\subsection{Piecewise Linear Oracle}
\label{sec:LinearOracle}

Our setup is not limited to only continuous oracle functions like the ones discussed so far in the preceeding sections. Our method can also work with piecewise-defined functions like
\beq
\varphi(\mathbf x) = 
\left\{
\begin{array}{ll}
-x^{(2)}, \qquad {\rm for \ }x^{(1)}<0, \\
+x^{(1)}, \qquad {\rm for \ }x^{(1)}\ge 0.
\end{array}
\right.
\label{eq:oracleDisc}
\eeq

\begin{figure}[t]
    \centering
    \includegraphics[width=0.45\textwidth]{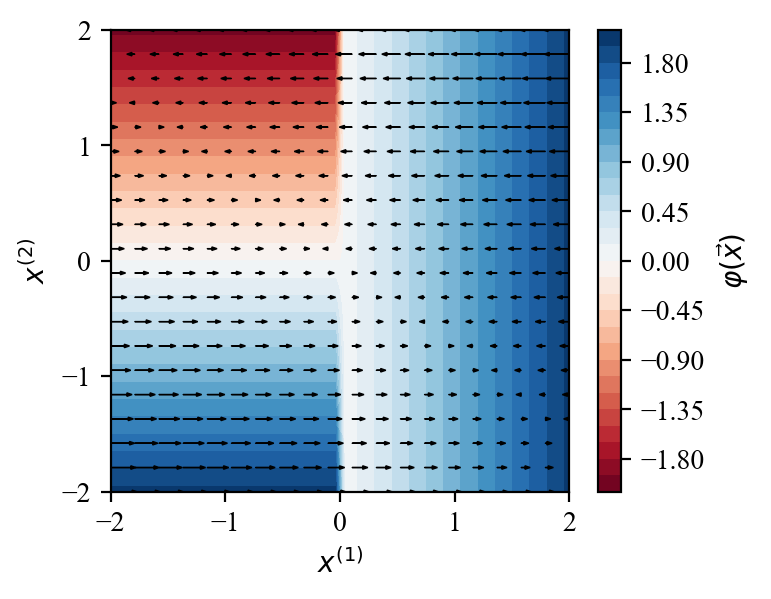}
    \includegraphics[width=0.45\textwidth]{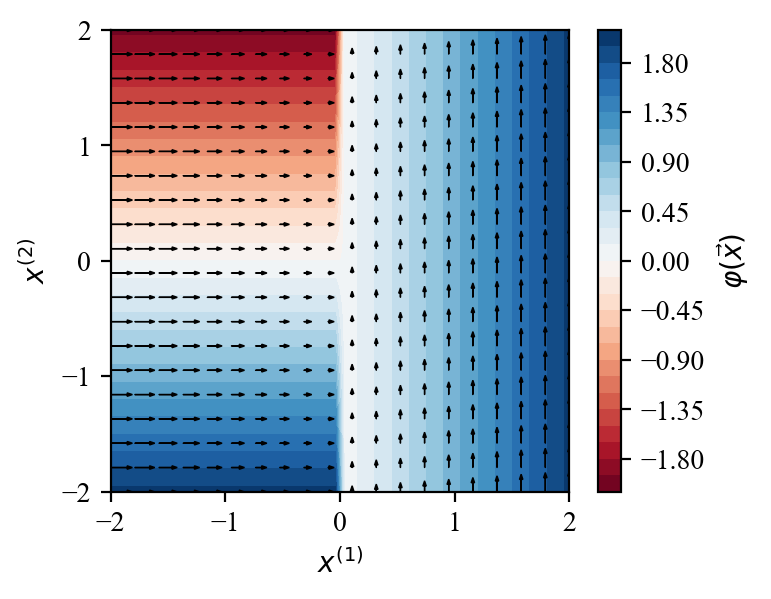}
    \caption{The symmetry generated by the oracle (\ref{eq:oracleDisc}) with a shallow method with no hidden layers and no bias (top panel) or a deep method with three hidden layers and bias (bottom panel). }\label{fig:Disc}
\end{figure}

The resulting oracle function is illustrated with the color map in Figure~\ref{fig:Disc}. Since the oracle is now a function which is not continuously differentiable, our parametrization of the symmetry transformation needs to be properly generalized.

The advantage of using a neural network as a universal function approximator is highlighted in Figure~\ref{fig:Disc}. In the top panel we use no hidden layers, while in the bottom panel we use a deep learning architecture with three hidden layers. The found symmetry transformation in each case is then shown with the vector field and superimposed on the oracle color map. The results are noticeably different, particularly near the locations of discontinuity in the oracle function. The deep-learning approach in the bottom panel correctly identifies a transformation which preserves the oracle values {\em everywhere} within the considered domain. In contrast, the shallow approach in the top panel is unable to adjust the transformation near the discontinuity, which leads to locations near the boundary $x^{(1)}=0$ where the arrows run across the equipotential contours, violating the conservation law.

\subsection{Manhattan Distance Oracle}
\label{sec:L1}

In this subsection we consider one more example of an oracle in two dimensions ($n=2$), which, while continuous, is not continuously differentiable:
\beq
\varphi(\mathbf x) = |x^{(1)}|+|x^{(2)}|.
\label{eq:oracleL1}
\eeq

\begin{figure}[t]
    \centering
    \includegraphics[width=0.45\textwidth]{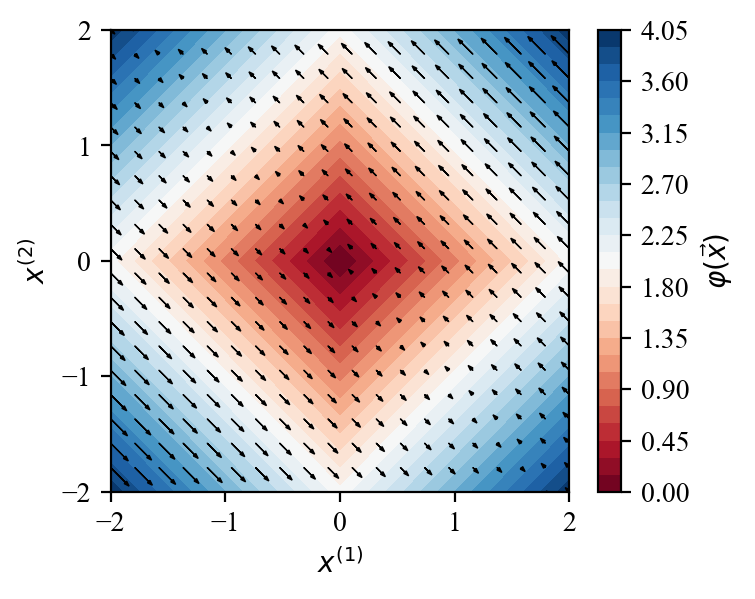}
    \includegraphics[width=0.45\textwidth]{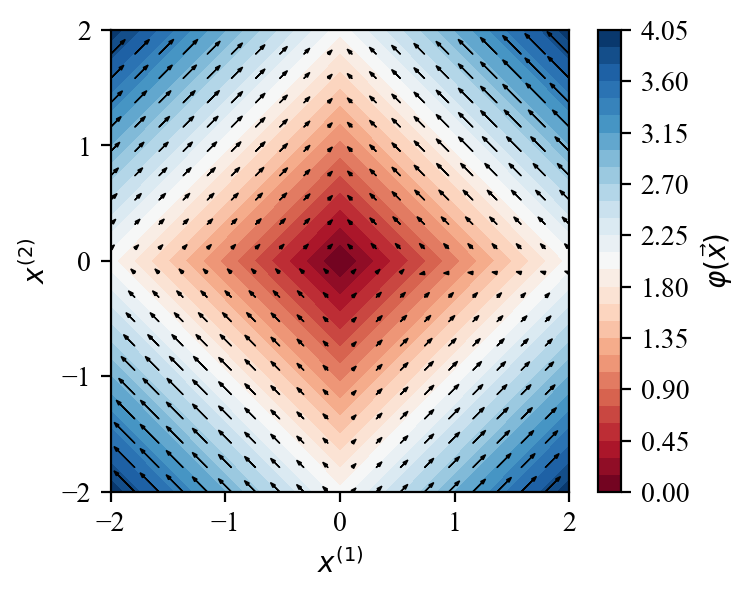}
    \caption{The symmetry generated by the L1 oracle (\ref{eq:oracleL1}) with a shallow method with no hidden layers and no bias (top panel) or a deep method with three hidden layers and bias (bottom panel). }\label{fig:L1}
\end{figure}

The results from our procedure are shown in Figure~\ref{fig:L1} in complete analogy to Figure~\ref{fig:Disc}. In the top panel we use no hidden layers, while in the bottom panel we use a deep learning architecture with three hidden layers and bias. We observe that the deep-learning approach can again correctly handle the discontinuities, always generating transformations along, but never across, the contours of equal oracle function values.

\section{Conclusions}
\label{sec:conclusions}

\begin{figure*}[t]
    \centering
    \includegraphics[width=0.90\textwidth]{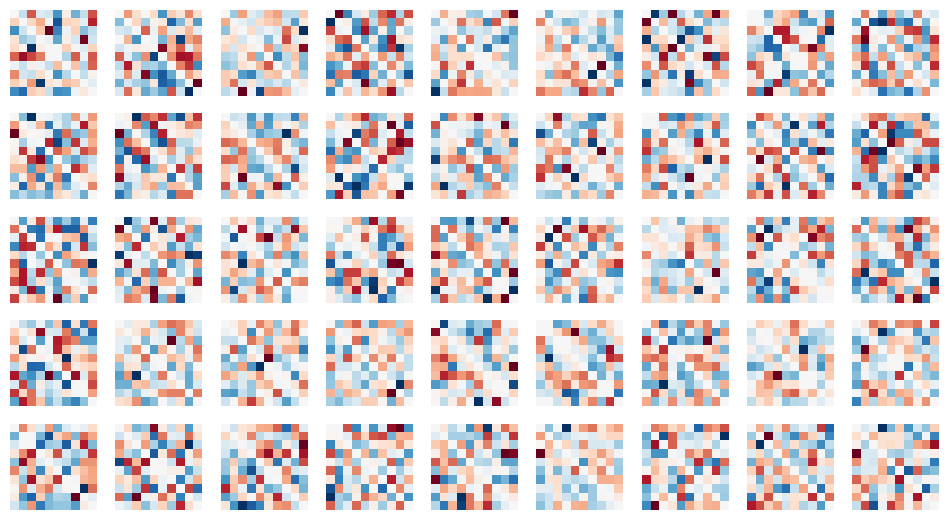}
    \caption{The set of $N_g=45$ generators found by our method in the case of the SO(10) group (length-preserving rotations in $n=10$ dimensions). Due to the complexity of the problem, the learning rate for this example was reduced to 0.01.}\label{fig:SO10}
\end{figure*}

In this paper, we studied a fundamental problem in data science which is commonly encountered in many fields: what is the symmetry of a labeled dataset, and how to identify its group structure? For this purpose, we designed a deep-learning method which models the generic symmetry transformation and its generators with a fully connected neural network. We then constructed suitable loss functions which ensure that the applied transformations are symmetries and that the corresponding set of generators forms a closed (sub)algebra. An important advantage of our approach is that we do not require any advance knowledge of what symmetries can be expected in the data, i.e., instead of testing for symmetries from a predefined list of possibilities, we learn the symmetry directly.

Our procedure is very general and is universally applicable in a wide variety of situations. The centerpiece of our analysis is an oracle $\varphi(\mathbf x)$ which defines the conserved quantity. The oracle can be completely general, as illustrated with several examples in the paper. For example, it can be a continuous bilinear function, as in the case of the Euclidean or Minkowski distances and the squeeze mapping; the corresponding symmetries were discussed in Sections~\ref{sec:rotations2dim}-\ref{sec:squeeze}. It can also be a discontinuous function (Section~\ref{sec:LinearOracle}) or a function which is not smooth and continuously differentiable (Section~\ref{sec:L1}).

In the process of deriving the full set of symmetries, the method also allows us to analyze the complete subgroup structure of the symmetry group. In the paper we worked out explicitly the subgroup structure of the rotation groups $SO(2)$ (Section~\ref{sec:rotations2dim}), $SO(3)$ (Section~\ref{sec:rotations3dim}) and $SO(4)$ (Section~\ref{sec:rotations4dim}), as well as the Lorentz group $SO(1,3)$ ((Section~\ref{sec:Lorentz})). As one last {\em tour de force} example, we successfully applied our method to the case of Euclidean length-preserving rotations in $n=10$ dimensions. The resulting 45 generators of the corresponding group $SO(10)$ commonly used in grand unification \cite{Gell-Mann:1979vob} are shown in Figure~\ref{fig:SO10}.

The symmetries discussed in this paper have important implications for simulation-based inference, and in particular parameter retrieval from observations. In a typical inverse problem scenario, the (possibly multi-dimensional) labels $y$ play the role of observed variables, and the features $\mathbf{x}$ play the role of input parameters. The parameter retrieval task is to infer $\mathbf{x}$ given $y$. In the presence of a symmetry this inversion is not unique, but results in a whole family of valid input parameters, all related by a symmetry. Therefore, knowing that there is a symmetry in the data to begin with can be very useful in parameter retrieval algorithms, especially when the data is very high dimensional and the symmetry is difficult to see by the human eye.

Note that the same approach can be extended or modified to solve other common problems and tasks in group theory and/or BSM model building not discussed in this paper.
\begin{itemize}
\item {\em Derivation of the Cartan subalgebra.} In our approach, this can be trivially accomplished by setting $a_{[\alpha\beta]\gamma}=0$ in (\ref{eq:closuremismatch}), which will force all the learned generators to commute. 
\item {\em Direct product decompositions.} We can also look for symmetries whose algebras can be expressed as direct sums of two separate subalgebras --- we just have to include closure terms in the loss function for each of the two individual subalgebras. 
\item {\em Internal symmetries.} Note that our approach can be readily generalized to look for internal symmetries \cite{Craven:2021ems}, which can prove useful in model building in high energy theory. 
\item {\em Canonical basis for the generators.} Since our method is basis-independent, the generators are obtained in a generic basis, where the structure constants may be difficult to recognize. We could, however, aid the program in finding a canonical basis of generators by including a term in the loss function which encourages sparsity among the structure constants.
\end{itemize}

In this paper we focused on idealized examples with no noise or statistical fluctuations in the determination of the labels. It will be instructive to test our method on noisy datasets, including real experimental data. We postpone this study to a future publication.

In conclusion, our study opens the door for using a machine learning approach in the study of Lie groups and their properties and further bridges the fields of formal mathematics and theoretical computer science.

\acknowledgements
We thank A.~Davis, S.~Gleyzer, K.~Kong, S.~Mrenna, H.~Prosper and P.~Shyamsundar for useful discussions. We thank P.~Ramond for group theory insights and inspiration.  This work is supported in part by the U.S.~Department of Energy award number DE-SC0022148.

\bibliography{references}

\end{document}